\documentclass[a4paper,11pt]{article}
\usepackage{jheppub} 
\usepackage{lineno}
\usepackage{bbm}
\usepackage{amsmath}
\usepackage{amssymb}
\usepackage{braket}
\usepackage{bm}
\usepackage{caption}
\usepackage{xcolor}

\title{\boldmath The spectrum of defect ABJM theory}

\author{Charlotte Kristjansen\footnote{Corresponding author.},}
\author{Xin Qian}
\author{and Chenliang Su}
\affiliation{Niels Bohr International Academy, Niels Bohr Institute, Copenhagen University, \\
Blegdamsvej 17, DK-2100 Copenhagen \O, Denmark}

\emailAdd{kristjan@nbi.dk}

\abstract{
We determine the spectrum of quantum fluctuations in a $\frac{1}{2}$-BPS domain wall version of ABJM theory, thereby enabling  the perturbative exploration of the corresponding defect CFT. As expected, the spectrum reflects the  supersymmetry
of the model.
}

\begin{document}
\maketitle
\flushbottom

\section{Introduction}
\label{sec:intro}

Defects in supersymmetric gauge theories on one hand constitute probes which can reveal different phases of the theories and on the other hand provide ways of partially or completely breaking the supersymmetry in a controlled manner, allowing to test,
challenge and eventually enlarge the domains of validity of exact methods such as localization, integrability and bootstrap. A natural starting point for such studies are 1/2-BPS defects in the maximally supersymmetric gauge theory in four dimensions,
${\cal N}=4$ super Yang Mills theory, where an extensive analysis of supersymmetric boundary conditions has been performed~\cite{Gaiotto:2008sa,Gaiotto:2008ak}. Among 1/2-BPS defects one finds configurations which correspond to so-called Nahm poles. These are configurations where one or more scalar fields have vacuum expectation values which diverge as the defect is approached and
include the cases of 't Hooft lines~\cite{tHooft:1977nqb,Kapustin:2005py}, Gukov-Witten surface defects~\cite{Gukov:2006jk}, domain walls~\cite{Diaconescu:1996rk,Constable:1999ac,Kristjansen:2012tn} and others~\cite{deLeeuw:2024qki}. The domain wall dual to the D3-D5 probe brane set-up with flux~\cite{Constable:1999ac} has turned out to be amenable both to supersymmetric localization
where one-point functions of chiral primaries have been computed exactly~\cite{Wang:2020seq,Komatsu:2020sup}  and to integrability where one-point functions of non-protected operators have been computed in an 
asymptotic limit~\cite{deLeeuw:2015hxa,Buhl-Mortensen:2015gfd,Buhl-Mortensen:2017ind,Gombor:2020kgu,Komatsu:2020sup,Kristjansen:2020mhn,Gombor:2020auk}. 
Furthermore, the perturbative program for  this defect configuration has been completed~\cite{Buhl-Mortensen:2016pxs,Buhl-Mortensen:2016jqo}, and the initial steps in the implementation of the boundary conformal program have been taken~\cite{deLeeuw:2017dkd,Baerman:2024tql}. The 't Hooft loop and the Gukov-Witten defects are likewise amenable to localization~\cite{Okuyama:2006jc,Choi:2024ktc}, the 't Hooft line indirectly via the treatment of the S-dual Wilson line.
Furthermore,  tools of integrability have made possible the computation of one-point functions at leading order for 
the 't Hooft line ~\cite{Kristjansen:2023ysz}, and there exists an asymptotic one-point formula valid up to a possible
dressing phase~\cite{Gombor:2024api}. Finally, the perturbative program has been completed in this
case as well allowing for the loop wise computation of more general local as well as non-local observables~\cite{Kristjansen:2023ysz,Kristjansen:2024map}.

With the above successes in mind an obvious open question is whether similar achievements can be obtained for three-dimensional ABJM theory~\cite{Aharony:2008ug} which is somewhat similar to  ${\cal N}=4$ SYM in the sense that it is a supersymmetric gauge theory with a well-understood string theory dual and has an underlying integrable spin chain describing its conformal operators~\cite{Minahan:2008hf}.  In ABJM theory there likewise exists a 1/2-BPS domain wall configuration where the classical values of some scalar fields diverge at the defect~\cite{Terashima:2008sy}. While the BPS equations determining the classical fields in the case of ${\cal N}=4$
SYM theory are the Nahm equations, in the case of ABJM theory they take the form of the so-called
Basu-Harvey equations~\cite{Basu:2004ed} whose solutions can be viewed as square roots of  Nahm solutions as  explained in~\cite{Kristjansen:2021abc} and recalled below.  The ABJM domain wall has been shown to be integrable at tree level meaning that all one-point functions can 
be computed in closed form~\cite{Gombor:2021hmj,Kristjansen:2021abc,Gombor:2022aqj}.  Other integrable boundary states have been identified 
in~\cite{Yang:2021hrl,Bai:2024qtg}.
A string theory analysis points towards the ABJM domain wall being integrable also in the strong coupling limit~\cite{Dekel:2011ja,Linardopoulos:2021rfq,Linardopoulos:2022wol}. At the moment the extension to the asymptotic regime in the spirit of~\cite{Gombor:2020auk} constitutes an
open problem. 
In the present paper our aim is to set up the perturbative program for ABJM theory with the 1/2 BPS domain
wall in order to enable the computation of other local and non-local observables. 
The crucial part of this endeavour is the diagonalization of a complicated quadratic action caused by the non-vanishing vacuum expectation values of the scalars.  Compared to the case of ${\cal N}=4$ SYM theory we face several new challenges. First,
since the present gauge theory is a Chern Simons theory the mixing between fields involves mass terms as well as kinetic
terms and all gauge field components take part in the mixing.
Secondly, in the case of ${\cal N}=4$ SYM the vacuum expectation values of the scalar fields were given directly in terms of $\mathfrak{su}(2)$ generators and one could almost immediately apply $\mathfrak{su}(2)$ representation theory to perform the relevant 
diagonalization~\cite{Buhl-Mortensen:2016pxs,Buhl-Mortensen:2016jqo}. In the present case the vevs are rather square roots of 
$\mathfrak{su}(2)$ generators and although representation 
theory of $\mathfrak{su}(2)$ is obviously again the key to the solution, new tricks are needed to proceed. 

Our paper is structured as follows. We start by reviewing the classical solution of ABJM theory that describes a 1/2
BPS domain wall in section~\ref{set-up}. Subsequently, we present  and analyze the quantum action which results from expanding around the classical solution. In particular, we outline our strategy for determining the spectrum of quantum
fluctuations.
A first step is a classification of the fields into easy and complicated fields with the terminology referring to the complexity of their mixing. We will then proceed with disentangling the mixing of the fields, treating the easy fields in section~\ref{easy field}
and the complicated fields in section~\ref{complicated field}. In the diagonalization procedure we derive and make use of a particular variant of fuzzy spherical harmonics which form a basis for rectangular matrices. Some technicalities in the derivation of these fuzzy spherical harmonics are relegated to appendix~\ref{Mfuzzy} while appendix \ref{AppendixCS}  is devoted to the solution of general Chern-Simons equations of motion.
Finally, in  section~\ref{conclusion} we summarize the spectrum, expose its supersymmetric structure and
make some concluding remarks.

\section{The field theory set-up}
\label{set-up}
\subsection{The domain wall as a BPS solution}
\label{BPS equation}
We start from the action of the ABJM theory which reads \cite{Minahan:2009te,Bandres_2008} 
\begin{equation}\label{ABJM action}
\begin{aligned}
\mathcal{L}=  \frac{k}{4 \pi} \operatorname{tr}&\left[\varepsilon^{\mu \nu \lambda}\left(-A_\mu \partial_\nu A_\lambda-\frac{2i}{3} A_\mu A_\nu A_\lambda+\hat{A}_\mu \partial_\nu \hat{A}_\lambda+\frac{2i}{3} \hat{A}_\mu \hat{A}_\nu \hat{A}_\lambda\right)\right. \\
& +D_\mu Y_A^{\dagger} D^\mu Y^A+\frac{1}{12} Y^A Y_A^{\dagger} Y^B Y_B^{\dagger} Y^C Y_C^{\dagger}+\frac{1}{12} Y^A Y_B^{\dagger} Y^B Y_C^{\dagger} Y^C Y_A^{\dagger} \\
& -\frac{1}{2} Y^A Y_A^{\dagger} Y^B Y_C^{\dagger} Y^C Y_B^{\dagger}+\frac{1}{3} Y^A Y_B^{\dagger} Y^C Y_A^{\dagger} Y^B Y_C^{\dagger}\\
&+i\bar{\psi}^{A\dagger}\gamma^{\mu}D_{\mu}\psi_A -\frac{1}{2}Y_A^{\dagger}Y^A\bar{\psi}^{B\dagger}\psi_B+\frac{1}{2}\bar{\psi}^{A\dagger}Y^B Y_B^{\dagger}\psi_A+Y_A^{\dagger}Y^B\bar{\psi}^{A\dagger}\psi_B\\
&\left.-\bar{\psi}^{A\dagger} Y^B Y_A^{\dagger} \psi_B
			 +\frac{i}{2}\epsilon^{ABCD}Y_A^{\dagger}\bar{\psi}_{B}Y_C^{\dagger} \psi_D-\frac{i}{2}\epsilon_{A B C D}Y^A\bar{\psi}^{B\dagger} Y^C \psi^{D\dagger}\right] ,
\end{aligned}
\end{equation}
where $A=1,2,3,4$ are $SU(4)$ R-symmetry indices and 
where we choose the Dirac representation $\gamma^{\mu}=(\sigma^3,i\sigma^1,i\sigma^2)$ with $\eta_{\mu\nu}=\mathrm{diag}(+,-,-)$. We follow the convention of \cite{Bandres_2008}, where $\dagger$ denotes complex conjugate plus a transpose on (and only on) the gauge index and $\bar{\psi}$ implies a transpose on the spinor index and right multiplication by $\gamma^0$ (without extra complex conjugate)\footnote{Although this notation is not very commonly used, it produces the usual fermionic action. For example, we have $i\bar{\psi}^{A\dagger}\gamma^{\mu}D_{\mu}\psi_A=i\bar{\psi}^{A*T_s T_g }\gamma^0\gamma^{\mu}D_{\mu}\psi_A$, where $*$ denotes the complex conjugate and $T_s(T_g)$ denotes the transpose on spinor(gauge) index.}. Also note that the transformation property of
a field under R-symmetry is indicated by an upper or lower $SU(4)$ index, while $\dagger$ or the bar notation has nothing to do with it. The covariant derivative is defined as 
\begin{equation}
    D_{\mu}X=\partial_{\mu}X + i A_{\mu}X -iX\hat{A}_{\mu},\ D_{\mu}X^{\dagger}=\partial_{\mu}X^{\dagger} + i \hat{A}_{\mu}X^{\dagger}- iX^{\dagger}A_{\mu}.
\end{equation}
We use the notation $x^\mu=(x^0,x^1,z)$ and consider a domain wall placed at $z=0$. The coordinates on the domain
wall are denoted as $x^a$ with $a=0,1$.  On one  side of the wall, $z<0$, all fields have vanishing vacuum expectation values whereas for $z>0$ two of the four scalar fields, $Y^{1}, Y^2 $,  have non-vanishing vevs. These vevs fulfil not only the classical
equations of motion but in addition a set of BPS equations which ensures that half of the supersymmetries are 
preserved~\cite{Terashima:2008sy,Kristjansen:2021abc}, i.e.\

\begin{equation}\label{BPS eq}
    \frac{dY_{\text{cl}}^{\alpha}}{dz}=\frac{1}{2}Y_{\text{cl}}^{\alpha}Y_{\text{cl},\beta}^{\dagger}Y_{\text{cl}}^{\beta}-\frac{1}{2}Y_{\text{cl}}^{\beta}Y_{\text{cl},\beta}^{\dagger}Y_{\text{cl}}^{\alpha}, \hspace{0,5cm} \alpha=1,2.
\end{equation}
A scale-invariant solution is found in \cite{Terashima:2008sy} 
\begin{equation}\label{BPS sol}
    Y_{\text{cl}}^{\alpha}=\frac{y^{\alpha}}{\sqrt{z}}, \quad z>0,
\end{equation}
where $y^{\alpha}$ is a $N\times N$ matrix but only the $(q-1)\times q$ upper left block has non-vanishing entries, where $q>1$. Precisely,
\begin{equation}\label{classical sols}
    y^1_{ij}=\delta_{i,j-1}\sqrt{i}, \quad y^2_{ij}=\delta_{ij}\sqrt{q-i}, \quad i=1,\dots,q-1, \quad j=1,\dots,q.
\end{equation}
These matrices  also appear in the study of the Coulomb branch of the mass deformed ABJM theory
\cite{Gomis_2008}.
The classical solution breaks the gauge symmetry down to $U(N-q+1)\times U(N-q)$ for finite $z$ and the global $R$ symmetry to $SU(2)\times SU(2)\times U(1)$.  The latter was made very explicit in the analysis of the classical fields
in~\cite{Kristjansen:2021abc} which we briefly summarize below.
Later we explicitly demonstrate that the quantum fields  likewise organize themselves
into irreducible representations of $\mathfrak{su}(2)$.

Defining 
\begin{equation}
    \Phi^{\alpha}_{\ \beta}=y^{\alpha}y_{\beta}^{\dagger},
\end{equation}
one can derive the following commutation relations~\cite{Kristjansen:2021abc}
\begin{equation}
    [\Phi^{\alpha}_{\ \beta},\Phi^{\gamma}_{\ \theta}]=\delta^{\alpha}_{\ \theta}\Phi^{\beta}_{\ \gamma}-\delta^{\beta}_{\ \gamma}\Phi^{\alpha}_{\ \theta},
\end{equation}
which is nothing but the  commutation relation for $U(2)$.  Forming the linear combination
\begin{equation}
\label{2.10}
    t_i=\frac{1}{2}(\sigma_i)^{\alpha}_{\ \beta}\Phi^{\beta}_{\ \alpha},
\end{equation}
one can show that  
\begin{equation}
    [t_i, t_j]=i\varepsilon_{ijk}t_k,
\end{equation}
which means that the $t_i$ are $N\times N$ matrices with $\mathfrak{su}(2)$ generators of size $(q-1)\times (q-1)$ in their 
upper left hand corner.
One can also define
\begin{equation}
    \hat{\Phi}_{\alpha}^{\ \beta}=y_{\alpha}^{\dagger}y^{\beta},
\end{equation}
and find the similar commutation relations
\begin{equation}
    [\hat{\Phi}_{\alpha}^{\ \beta},\hat{\Phi}_{\gamma}^{\ \theta}]=-\delta_{\alpha}^{\ \theta}\hat{\Phi}_{\gamma}^{\ \beta}+\delta^{\beta}_{\ \gamma}\hat{\Phi}_{\alpha}^{\ \theta},
\end{equation}
Then the matrices $\hat{t_i}$ defined by
\begin{equation}
    \hat{t}_i=-\frac{1}{2}(\sigma_i)_{\alpha}^{\ \beta}\hat{\Phi}_{\beta}^{\ \alpha},
\end{equation} contain a $q$-dimensional representation of $\mathfrak{su}(2)$ in their upper left hand corner, since
\begin{equation}
    [\hat{t}_i,\hat{t}_j]=i\varepsilon_{ijk}\hat{t}_k.
\end{equation}
 One can also derive the following useful identities
\begin{equation}
\label{2.16}
    y^{\alpha}y_{\alpha}^{\dagger}=q\mathbbm{1}_{q-1}, \quad y_{\alpha}^{\dagger}y^{\alpha}=(q-1)\mathbbm{1}_{q}.
\end{equation}

\subsection{The perturbative action}\label{expanded action}
We expand the action around the classical configuration, i.e.\ we make the following replacement
\begin{equation*}
    Y^{\alpha}\rightarrow y^{\alpha}+\widetilde{Y}^{\alpha},
\end{equation*}
where $y^{\alpha}$ is the classical solution given in eq.\ (\ref{BPS sol}) and $\widetilde{Y}^{\alpha}$ the corresponding quantum fluctuation.
For simplicity we will leave out the tilde on the quantum field in the following.
The expanded action then reads
\begin{equation}
    \mathcal{L}=\mathcal{L}_{\text{b}}+\mathcal{L}_{\text{f}}+\mathcal{L}_{\text{int}},
\end{equation}
where $\mathcal{L}_b$ and $\mathcal{L}_f$ are Gaussian terms  for bosonic and fermionic fields respectively and 
where $\mathcal{L}_{\text{int}}$ are interaction terms. The linear terms vanish by the BPS conditions.
The non-vanishing vevs introduce a highly non-trivial mixing between
both flavour and colour components of the various fields. Our aim will be to disentangle this
mixing and find the spectrum of the quantum fluctuations which is a necessary prerequisite
for enabling perturbative computations. 
We do not add gauge fixing terms at the present stage but fix the gauge by
imposing constraints on the solutions later on.

 The bosonic Gaussian action can be organized as
\begin{equation}
		\begin{aligned}
			\mathcal{L}_{\text{b}}=\frac{k}{4\pi}\text{tr}\bigg \{&-\epsilon^{\mu\rho\nu}A_{\mu}\partial_{\rho}A_{\nu} + \epsilon^{\mu\rho\nu}\hat{A}_{\mu}\partial_{\rho}\hat{A}_{\nu}+\left( A-\hat{A}\ \text{mixing} \right)\\
			&+\left( \partial_{\mu}Y^{\alpha} \right)\partial^{\mu}Y_{\alpha}^{\dagger}+\left( Y^{\dagger}-Y\ \text{mixing} \right)+\left( Y-A\ \text{mixing} \right)\bigg\},
		\end{aligned}
	\end{equation}
where the mixing terms involving gauge fields are given by
\begin{equation}
    \left( A-\hat{A}\ \text{mixing} \right)=-\frac{2}{z}\hat{A}_{\mu}y_{\alpha}^{\dagger}A^{\mu}y^{\alpha}+\frac{1}{z}y_{\alpha}^{\dagger}A_{\mu}A^{\mu}y^{\alpha}+\frac{1}{z}\hat{A}_{\mu}y_{\alpha}^{\dagger}y^{\alpha}\hat{A}^{\mu},
\end{equation}
as well as
\begin{equation}\label{YAmixing}
\begin{aligned}
    \left( Y-A\ \text{mixing} \right)=2i&\left(\partial_{\mu}\frac{y^{\alpha}}{\sqrt{z}}\right)\left(\hat{A}^{\mu}Y_{\alpha}^{\dagger}-Y_{\alpha}^{\dagger}A^{\mu}\right)+2i\left(\partial_{\mu}\frac{y_{\alpha}^{\dagger}}{\sqrt{z}} \right)\left(A^{\mu}Y^{\alpha}-Y^{\alpha}\hat{A}_{\mu}\right)\\
    &+\frac{i}{z^{1/2}}(\partial_{\mu}A^{\mu})\left( Y^{\alpha}y_{\alpha}^{\dagger}-y^{\alpha}Y_{\alpha}^{\dagger} \right)+\frac{i}{z^{1/2}}(\partial_{\mu}\hat{A}^{\mu})\left( Y_{\alpha}^{\dagger}y^{\alpha}-y_{\alpha}^{\dagger}Y^{\alpha} \right).
    \end{aligned}
\end{equation}
For the scalar fields there is no mixing in the kinetic terms but the mass-like terms can be naturally split in the following
way according to the mixing pattern
\begin{equation}
    \left( Y^{\dagger}-Y\ \text{mixing} \right)= m^2_{Y^{\alpha}Y_{\alpha}^{\dagger}}+m^2_{Y^{\alpha}Y_{\beta}^{\dagger}}+m^2_{YY},
\end{equation}
where 
\begin{equation}
		\begin{aligned}\label{YAYA}
			m^2_{Y^{A}Y_{A}^{\dagger}}=&\frac{1}{4z^2}Y^{A} Y_{A}^{\dagger} y^{\beta} y_{\beta}^{\dagger} y^{\gamma} y_{\gamma}^{\dagger}
			-\frac{1}{2z^2}Y^{A} Y_{A}^{\dagger} y^{\beta} y_{\gamma}^{\dagger} y^{\gamma} y_{\beta}^{\dagger}
			-\frac{1}{2z^2}Y^{A} y_{\beta}^{\dagger} y^{\beta} Y_{A}^{\dagger} y^{\gamma} y_{\gamma}^{\dagger}\\
			&+\frac{1}{z^2}Y^{A} y_{\beta}^{\dagger} y^{\gamma} Y_{A}^{\dagger} y^{\beta} y_{\gamma}^{\dagger}+\frac{1}{4z^2}Y_{A}^{\dagger} Y^{A} y_{\beta}^{\dagger} y^{\beta} y_{\gamma}^{\dagger} y^{\gamma}-\frac{1}{2z^2}Y_{A}^{\dagger}Y^{A}   y_{\beta}^{\dagger} y^{\gamma} y_{\gamma}^{\dagger} y^{\beta}, \\
		\end{aligned}
	\end{equation}
 
\begin{equation}
		\begin{aligned} \label{YAYAD}
			m^2_{Y^{\alpha}Y_{\beta}^{\dagger}}=& \frac{1}{4z^2}\left(Y^{\alpha} y_{{\alpha}}^{\dagger} y^{\beta} Y_{\beta}^{\dagger} y^{\gamma} y_{{\gamma}}^{\dagger}+y_{\alpha}^{\dagger}Y^{\alpha}Y_{\beta}^{\dagger}y^{\beta}y_{\gamma}^{\dagger}y^{\gamma}+Y^{\alpha} y_{{\alpha}}^{\dagger} y^{\beta} y_{{\beta}}^{\dagger} y^{\gamma} Y_{\gamma}^{\dagger}+Y^{\alpha} y_{\beta}^{\dagger} y^{\beta} Y_{\gamma}^{\dagger} y^{\gamma} y_{{\alpha}}^{\dagger}\right) \\
			 -&\frac{1}{2z^2}\left(Y^{\alpha} Y_{\beta}^{\dagger} y^{\beta} y_{{\alpha}}^{\dagger} y^{\gamma} y_{{\gamma}}^{\dagger}+Y^{\alpha} y_{\beta}^{\dagger} y^{\beta} y_{{\alpha}}^{\dagger} y^{\gamma} Y_{\gamma}^{\dagger}+y^{\alpha} y_{\beta}^{\dagger} Y^{\beta} Y_{\alpha}^{\dagger} y^{\gamma} y_{{\gamma}}^{\dagger}+y^{\alpha} y_{\beta}^{\dagger} y^{\beta} Y_{\alpha}^{\dagger} Y^{\gamma} y_{{\gamma}}^{\dagger}\right. \\
			& \left.+y^{\alpha} y_{\beta}^{\dagger} Y^{\beta} y_{{\alpha}}^{\dagger} y^{\gamma} Y_{\gamma}^{\dagger}+y^{\alpha} Y_{\beta}^{\dagger} y^{\beta} y_{{\alpha}}^{\dagger} Y^{\gamma} y_{{\gamma}}^{\dagger}\right)+\frac{1}{z^2}\left(Y^{\alpha} Y_{\beta}^{\dagger} y^{\gamma} y_{{\alpha}}^{\dagger} y^{\beta} y_{{\gamma}}^{\dagger}+Y^{\alpha} y_{\beta}^{\dagger} y^{\gamma} y_{{\alpha}}^{\dagger} y^{\beta} Y_{\gamma}^{\dagger}\right),
		\end{aligned}
	\end{equation}

\begin{equation}
    \begin{aligned} \label{YY}
     	m^2_{YY}=	& \frac{1}{4z^2}\left(Y^{\alpha} y_{{\alpha}}^{\dagger} Y^{\beta} y_{\beta}^{\dagger} y^{\gamma} y_{{\gamma}}^{\dagger}+Y^{\alpha} y_{\beta}^{\dagger} Y^{\beta} y_{{\gamma}}^{\dagger} y^{\gamma} y_{{\alpha}}^{\dagger}+y^{\alpha} Y_{\alpha}^{\dagger} y^{\beta} Y_{\beta}^{\dagger} y^{\gamma} y_{{\gamma}}^{\dagger}+y^{\alpha} Y_{\beta}^{\dagger} y^{\beta} Y_{\gamma}^{\dagger} y^{\gamma} y_{{\alpha}}^{\dagger}\right) \\
     		& -\frac{1}{2z^2}\left(Y^{\alpha} y_{\beta}^{\dagger} Y^{\beta} y_{{\alpha}}^{\dagger} y^{\gamma} y_{{\gamma}}^{\dagger}+Y^{\alpha} y_{\beta}^{\dagger} y^{\beta} y_{{\alpha}}^{\dagger} Y^{\gamma} y_{{\gamma}}^{\dagger}+y^{\alpha} Y_{\beta}^{\dagger} y^{\beta} Y_{\alpha}^{\dagger} y^{\gamma} y_{{\gamma}}^{\dagger}+y^{\alpha} y_{\beta}^{\dagger} y^{\beta} Y_{\alpha}^{\dagger} y^{\gamma} Y_{{\gamma}}^{\dagger}\right. \\
     		& \left.+y^{\alpha} y_{\beta}^{\dagger} Y^{\beta} y_{{\alpha}}^{\dagger} Y^{\gamma} y_{{\gamma}}^{\dagger}+y^{\alpha} Y_{\beta}^{\dagger} y^{\beta} y_{\alpha}^{\dagger} y^{\gamma} Y_{\gamma}^{\dagger}\right)+\frac{1}{z^2}\left(Y^{\alpha} y_{\beta}^{\dagger} Y^{\gamma} y_{{\alpha}}^{\dagger} y^{\beta} y_{{\gamma}}^{\dagger}+y^{\alpha} Y_{\beta}^{\dagger} y^{\gamma} Y_{\alpha}^{\dagger} y^{\beta} y_{{\gamma}}^{\dagger}\right).
     	\end{aligned}
\end{equation}
The subscripts indicate the type of coupling between the scalar fields in the various terms. We notice that 
the terms in eq.~(\ref{YAYA}) do not involve any flavour mixing whereas the terms in eq.~(\ref{YAYAD}) and eq.~(\ref{YY}) do.

The ABJM action is already quadratic in the fermionic fields $\psi$, hence the perturbative Gaussian action $\mathcal{L}_{\text{f}}$ is obtained by simply replacing the scalar fields $Y$ in the fermionic part of~(\ref{ABJM action}) by their classical solutions, and we do not repeat the expression here. 

\subsection{On the spectrum of the theory\label{Exp-spectrum}}
The expanded action above exhibits a very involved
mixing between both flavour and color components of 
fields. For certain field components there appear mass-like terms  
 which depend on the coordinate transverse to the defect as
$m^2/x_d^2$ for bosons and as $m_f/x_d$ for fermions. 
Our aim is to diagonalize the quadratic part of
the action and thus prepare the theory for perturbative computations. The consequence of the space-dependent 
masses for certain field components is well-known. The field components in question are confined to moving in 
a half space, here $x_d=x_2\equiv z>0$,  and their propagators become propagators in an auxiliary $AdS_{d+1}$ space which has the transverse coordinate, $x_d$, as its radius and the domain wall as its boundary~\cite{Buhl-Mortensen:2016pxs,Buhl-Mortensen:2016jqo}. 
 Propagators in an $AdS_{d+1}$ space can be expressed in terms of modified Bessel functions or equivalently
 in terms of  hypergeometric functions. More specifically, a scalar propagator $G({\bf{x}},{\bf{y}})$ which fulfils the equation
 below in $(d+1)$-dimensional Euclidean space
 \begin{equation}
\left( -\partial_\mu \partial^\mu+ \frac{m^2}{x_d^2}\right)G({\bf{x}},{\bf{y}})  =\delta ({\bf{x}}-{\bf{y}}),
 \end{equation}
 can by a Weyl transformation be related to a propagator in $AdS_{d+1}$  in the following way
 \begin{equation}
 G({\bf{x}},{\bf{y}})= \frac{G_{\mbox{\footnotesize{$AdS$}}}\,({\bf{x}},{\bf{y}})}{(x_d \,y_d)^{\frac{d-1}{2}}}, 
 \end{equation}
where the AdS mass is given by
\begin{equation}
m^2_{\footnotesize \mbox{$AdS$}} = m^2-\frac{d^2-1}{4}.
\end{equation}
Instead of characterizing a field in terms of its mass it is convenient to use the parameter $\nu$ defined as
\begin{equation}
\nu=\sqrt{m^2+\frac{1}{4}}.
\end{equation}
In order for a theory in  $AdS_{d+1}$ space to be stable its scalar excitations have to obey the BF bound~\cite{Breitenlohner:1982jf}
\begin{equation}
m^2_{\footnotesize \mbox{$AdS$}} \geq -\frac{d^2}{4},
\end{equation}
which implies
\begin{equation}
m^2\geq -\frac{1}{4} \hspace{0.5cm} \mbox{or} \hspace{0.5cm} \nu \geq 0.
\end{equation}

The scalar propagator in Euclidean $AdS_{d+1}$ can be written in the following form~\cite{Liu:1998ty} 
\begin{equation}\label{eq:PropagatorBessel}
\begin{aligned}
G_{{AdS}}({\bf x},{\bf y})
&=(x_d \,y_d)^{d/2} \int \frac{d^d \vec{k}}{(2\pi)^d} \int _0^\infty d w \frac{w}{w^2+\vec{k}^2}e^{i \vec{k}\cdot(\vec{x}-\vec{y})} J_\nu (w x_d)J_\nu (w y_d),\\
&=(x_d \,y_d)^{d/2} \int \frac{d^d \vec{k}}{(2\pi)^d}e^{i \vec{k}\cdot(\vec{x}-\vec{y})} I_{\nu}(|\vec{k}| x_d^<) K_{\nu}(|\vec{k}| x_d^>),
\end{aligned} 
\end{equation}
where $I$ and $K$ are modified Bessel functions with $x_d^<$ ($x_d^>$) the smaller (larger) of $x_d$ and $y_d$.
The parameter $\nu$ describes the scaling behavior of the field near the boundary of AdS$_{d+1}$ and is related to the conformal dimension, $\hat{\Delta}$, of the dual operator in the CFT$_d$ which lives on the 
boundary.\footnote{The precise definition of the conformal boundary operators requires that one performs a rescaling of the field components and organize these
 into gauge invariant objects. We refer to~\cite{deLeeuw:2017dkd} for an example of how this procedure works for ${\cal N}=4$ SYM.}
More precisely~\cite{Witten:1998qj}
\begin{equation}
\hat{\Delta}= \nu+\frac{d}{2}.
\end{equation}
A fermionic field with a mass term of the type $m_f/x_d$ in $d$-dimensional Euclidean space translates into a fermion in Euclidean
$AdS_{d+1}$ with the same mass, and  the conformal dimension of the dual operator is given by
\begin{equation}
\hat{\Delta}= |m_f|+ \frac{d}{2}.
\end{equation}
The fermionic propagator in $AdS_{d+1}$ can be found from the scalar one as explained e.g.\ in~\cite{Kawano:1999au}.
 For a superconformal field theory operators organize into multiplets where the conformal dimensions of bosons and fermions differ by half integers.  As the domain wall we are considering is 1/2 BPS we expect this pattern of conformal dimensions to be
visible in the spectrum of its quantum fluctuations.  We shall demonstrate by explicit computation that this is indeed the case.

\subsection{Classifying the fields}
To prepare for the diagonalization, we perform a classification of our fields, first in flavor space and subsequently in color space. As for flavour space we notice
that the scalar fields $Y^{\alpha=3,4}$ do not appear, neither in eq.~(\ref{YAmixing}), nor in
eq.~(\ref{YAYAD}) and eq.~(\ref{YY}) due to the fact that the corresponding classical fields are vanishing.
Hence, these scalars do not mix with gauge field components  and also not with the scalars $Y^{\alpha=1,2}$. We shall
therefore denote  $Y^{\alpha=3,4}$ as easy bosons. As opposed to these the scalar fields $Y^{\alpha=1,2}$ mix 
with  the $z$-component of the gauge field via the terms in eq.\ (\ref{YAmixing}) and  with all components of the gauge field via the 
kinetic Chern-Simons terms.
 Hence, we separate the bosonic fields in the following two categories
\begin{flalign}
    &\quad\quad\quad\quad\text{easy bosons:} && Y^{\Tilde{\alpha}}, \quad \Tilde{\alpha}=3,4,&&\nonumber \\
    &\quad\quad\quad\quad\text{complicated bosons:} && Y^{\alpha},\ A_{\mu},\ \hat{A}_{\mu}, \quad  \alpha=1,2,\ \mu=0,1,z.&&\nonumber
\end{flalign}
In a standard gauge theory (as opposed to a Chern-Simons gauge theory) one would seek to eliminate mixing terms involving
derivatives of fields such as some of the terms in eq. (\ref{YAmixing}) by working in background field gauge. This gauge choice, however, does not considerably simplify the mixing problem as we are still left with the kinetic Chern-Simons terms. We shall therefore not add gauge fixing terms at the present stage but solve the equations of motions for the various fields and impose the appropriate gauge constraints on the solutions later. Solving the equations of motion suffices for determining the
scaling behavior of a field near the $AdS$ boundary and thus the conformal dimension of its dual operator.

In the same way we split the fermionic fields into easy and complicated according to the complexity of their mixing as follows (from now on unless  explicitly stated, $\alpha$ and $\tilde{\alpha}$ will denote flavor indices),
\begin{flalign}
    &\quad\quad\quad\quad\text{easy fermions:} &&\psi_{{\alpha}}, \quad  {\alpha}=1,2, &&\nonumber\\
    &\quad\quad\quad\quad\text{complicated fermions:} && \psi_{\tilde{\alpha}}, \quad \tilde{\alpha}=3,4.&&\nonumber
\end{flalign}
In color space it is convenient to perform a block decomposition of the quantum fields
\begin{equation}
		X=\left( \begin{aligned}
		\quad	X(\nwarrow)\quad\quad & X(\nearrow)\quad \\[2pt] \quad  X(\swarrow)\quad\quad & X(\searrow)\quad
		\end{aligned} \right),
	\end{equation}
where the sizes of the various blocks of a given $X$ is defined by the size of its upper left hand block which can be found in
the list below
\begin{itemize}
        \item $Y^A(\nwarrow)$ is of size $(q-1)\times q$,
        \item $A^{\mu}(\nwarrow)$ is of  size $(q-1)\times (q-1)$,
        \item $\hat{A}^{\mu}(\nwarrow)$ is of size $q\times q$,
        \item $\psi_A(\nwarrow)$ is of  size $(q-1)\times q$.
\end{itemize}
This splitting makes the breaking of the gauge symmetry manifest. For example, 
\begin{equation}
    \text{tr}\left(Y^{\alpha}Y_{\beta}^{\dagger}y^{\beta}y_{\alpha}^{\dagger}\right)=\text{tr}\left( Y^{\alpha}(\nwarrow)Y_{\beta}^{\dagger}(\nwarrow)y^{\beta}y_{\alpha}^{\dagger} \right)+\text{tr}\left( Y^{\alpha}(\nearrow)Y_{\beta}^{\dagger}(\swarrow)y^{\beta}y_{\alpha}^{\dagger} \right),
    \label{2.19}
\end{equation}
which highlights the fact that the unbroken gauge group is $U(N-q+1)\times U(N-q)$. Here and in the following, a classical field which appears in connection with specific block components of quantum fields should be understood 
as just the non-vanishing matrix part of the classical field.
Due to the matrix structure of the classical fields, $y^{\alpha}$, the block decomposition above is also useful for 
exposing which field components couple to each other. E.g. 
in eq. (\ref{2.19}), the diagonal ($\nwarrow$, $\searrow$)  and off-diagonal terms ($\nearrow$, $\swarrow$)decouple. We likewise have the following useful observation
\begin{equation}\label{block division formula}
    \begin{aligned}    
    \text{tr}\left(Y^{\alpha} y_{\beta}^{\dagger}Y^{\beta} y_{\alpha}^{\dagger}y^{\gamma} y_{\gamma}^{\dagger}\right)=\text{tr}\left(Y^{\alpha}(\nwarrow)y_{\beta}^{\dagger}Y^{\beta}(\nwarrow)y_{\alpha}^{\dagger}y^{\gamma}y_{\gamma}^{\dagger}\right),
    \end{aligned}
\end{equation}
with similar relations being valid  for other quantum fields. Comparing eq.\  (\ref{2.19}) and eq. (\ref{block division formula}) 
one notices 
that only the mixing terms where the quantum  fields are adjacent contribute to the mixing between off-diagonal blocks.

\section{Diagonalization of the easy fields}
\label{easy field}
In this section, we  diagonalize the mass matrix of the easy bosons and the easy fermions. The mixing can be completely
resolved by means of $\mathfrak{su}(2)$ representation theory. The fields which diagonalize the mass matrix fall into irreducible
representations of $\mathfrak{su}(2)$ and their masses are given in terms of the corresponding Casimir.

\subsection{The easy bosons}

We start with the Gaussian part of the action with only easy bosons (for simplicity we leave out the factor of $k/4\pi$ in the 
following)
\begin{equation}
    \begin{aligned}
        \mathcal{L}_{\text{b,e}}=\text{tr}\left\{ (\partial_{\mu}Y^{\Tilde{\alpha}})\partial^{\mu}Y_{\Tilde{\alpha}}^{\dagger}+m_{Y^{\Tilde{\alpha}}Y_{\Tilde{\alpha}}^{\dagger}} \right\},
    \end{aligned}
\end{equation}
where $m_{Y^{\Tilde{\alpha}}Y_{\Tilde{\alpha}}^{\dagger}}$ is obtained by replacing $Y^A$($Y_A^{\dagger}$) with $Y^{\Tilde{\alpha}}$($Y_{\Tilde{\alpha}}^{\dagger}$) in eq. (\ref{YAYA}). 

Using the expressions for the bilinear combinations of classical fields presented  in Section \ref{BPS equation}, we can rewrite the mixing as
\begin{equation} 
\text{tr}\left(m_{Y^{\Tilde{\alpha}}Y_{\Tilde{\alpha}}^{\dagger}}\right)=\frac{1}{z^2}\text{tr}\left( -Y_{\Tilde{\alpha}}^{\dagger}L_i L_i Y^{\Tilde{\alpha}} \right),
\end{equation}
where
\begin{equation}\label{def of Li}
    L_i X = t_i X + X \hat{t}_i^T.
\end{equation}
One finds the $L_i$ of the $\pi_{q-1}\otimes\pi_{q}$ representation of $\mathfrak{su}(2)$, where $\pi_d$ denotes the $d$-dimensional (in other words, spin $\ell=(d-1)/2$) irreducible representation. Then by the block decomposition introduced in Section \ref{expanded action}, we can  write
\begin{equation}
\begin{aligned}    \text{tr}\left(m_{Y^{\Tilde{\alpha}}Y_{\Tilde{\alpha}}^{\dagger}}\right)=-&\frac{1}{z^2}\text{tr}\left( Y_{\Tilde{\alpha}}^{\dagger}(\nwarrow)L_i L_i Y^{\Tilde{\alpha}}(\nwarrow) \right)\\
    -&\frac{1}{z^2} \text{tr}\left( Y_{\Tilde{\alpha}}^{\dagger}(\swarrow)t_i t_i Y^{\Tilde{\alpha}}(\nearrow) \right)\\
    -&\frac{1}{z^2} \text{tr}\left( Y_{\Tilde{\alpha}}^{\dagger}(\nearrow)Y^{\Tilde{\alpha}}(\swarrow)\hat{t}_i^T \hat{t}_i^T  \right).
    \end{aligned}
\end{equation}
Hence, we conclude that 
\begin{itemize}
		\item  $Y_{\Tilde{\alpha}}^{\dagger}(\nwarrow)$ and $Y^{\Tilde{\alpha}}(\nwarrow)$ are mixed in the $\pi_{q-1}\otimes\pi_{q}$ representation specified by  $\mathcal{C}_2^{(q-1)\otimes q}$,
		\item  $Y_{\Tilde{\alpha}}^{\dagger}(\swarrow)$ and $Y^{\Tilde{\alpha}}(\nearrow)$ are mixed in the $\pi_{q-1}$ representation specified by 
		 $\mathcal{C}_2^{q-1} $,
		\item $Y_{\Tilde{\alpha}}^{\dagger}(\nearrow)$ and $Y^{\Tilde{\alpha}}(\swarrow)$ are mixed in the $\pi_{q}$ representation specified by
		$  \mathcal{C}_2^q $,
	\end{itemize}
where $\mathcal{C}^d_2$ is the Casimir of $\pi_d$ and $\mathcal{C}^{d\otimes d'}$ is that of $\pi_{d}\otimes\pi_{d'}$.

While the off-diagonal blocks are hence automatically organized in  definite irreducible representations, the diagonal blocks are not  and one needs to decompose $\pi_{q-1}\otimes
\pi_q$ in the usual way
\begin{equation}
    \pi_q \cong\, \bigoplus_{r=1}^{q-1} \pi_{2r} = \bigoplus_{\boldsymbol{\ell}=1/2}^{q-3/2}\boldsymbol{\ell},
\end{equation}
where the bold letter $\boldsymbol{\ell}$ denotes the spin $\ell$ irreducible representation.  The explicit decomposition into eigenstates is provided by a variant of the usual fuzzy
spherical harmonics which provides a basis for rectangular matrices as opposed to square matrices~\cite{Grosse:1995jt,Baez:1998he,Dasgupta:2002hx,Nastase:2009ny,Adachi:2020asg}.  For the present problem we need modified fuzzy spherical
harmonics of size $q(q-1)$, and in appendix~\ref{Mfuzzy} we demonstrate explicitly how one can construct these combining the  standard fuzzy spherical harmonics with the classical
fields.

Then, following the strategy of~\cite{Buhl-Mortensen:2016jqo}, the diagonalization of the $(\nwarrow)$-blocks can be done by 
expanding the fields in the basis consisting of these modified  fuzzy spherical harmonics
\begin{equation}
    Y^{\tilde{\alpha}}=Y_{\ell,m}^{\tilde{\alpha}} T_{\ell+1/2}^{m+1/2},
\end{equation}
where there is an implicit summation over repeated indices (e.g. $\ell$ and $m$ here) and where the $T$'s fulfil
\begin{equation}\label{L3 on T}
    L_3 T_{\ell+1/2}^{m+1/2} = (m+\frac{1}{2}) T_{\ell+1/2}^{m+1/2}, \quad L_i L_i T_{\ell+1/2}^{m+1/2} = (\ell+\frac{1}{2})(\ell+\frac{3}{2})T_{\ell+1/2}^{m+1/2},
\end{equation}
with $\ell=0,1,\dots,q-2$ and $m=-\ell-1,-\ell,\dots
,\ell$. Up to a normalization factor, the modified fuzzy spherical harmonics can be expressed in terms of the usual fuzzy spherical harmonics, $\hat{Y}_{\ell}^{m}$, and the 
classical fields $y^{1,2}$ in the following way
\begin{equation}
    T_{\ell+1/2}^{m+1/2}=-\sqrt{l-m}\,y^1\hat{Y}_{\ell}^{m+1}+\sqrt{\ell+m+1}\,y^2\hat{Y}_{\ell}^{m}.
\end{equation}
For details we refer to Appendix \ref{Mfuzzy}.

Now, the mixing terms of the easy bosons can be written as
\begin{equation}
\begin{aligned}    \text{tr}\left(m_{Y^{\Tilde{\alpha}}Y_{\Tilde{\alpha}}^{\dagger}}\right)=-&\frac{1}{z^2}(\ell+\frac{1}{2})(\ell+\frac{3}{2})Y_{\tilde{\alpha};\ell,m}^* Y^{\tilde{\alpha}}_{\ell,m}\\
-&\frac{1}{z^2}\frac{q(q-2)}{4}Y_{\tilde{\alpha};n\tilde{a}}^*Y^{\tilde{\alpha}}_{n\tilde{a}}\\
-&\frac{1}{z^2}\frac{(q-1)(q+1)}{4}Y_{\tilde{\alpha};a\tilde{n}}^* Y^{\tilde{\alpha}}_{a\tilde{n}}.
    \end{aligned}
\end{equation}
where $n=1,2,\dots,q-1$, $\tilde{n}=1,2,\dots
,q$, $a=q,q+1,\dots,N$ and $\tilde{a}=q+1,q+2,\dots,N$. Finally,  from the above equation one reads off the spectrum of easy bosons, which is shown in Table \ref{tab spec EB}.

\begin{table}[htbp]
\centering
\begin{tabular}{c|c}
\multicolumn{2}{c}{$Y^{\Tilde{\alpha}}(\nwarrow)$} 
\\
\hline
mass$^2$&multiplicity\\
\hline
$\ell(\ell+1)$ & $2(2\ell+1)$\\
\hline
\end{tabular}\\[10pt] 
\begin{tabular}{c|c}
\multicolumn{2}{c}{$Y^{\Tilde{\alpha}}(\nearrow)$} 
\\
\hline
mass$^2$&multiplicity\\
\hline
$(q-2)q/4$ & $2(q-1)(N-q)$\\
\hline
\end{tabular}\\[10pt]
\begin{tabular}{c|c}
\multicolumn{2}{c}{$Y^{\Tilde{\alpha}}(\swarrow)$} 
\\
\hline
mass$^2$&multiplicity\\
\hline
$(q-1)(q+1)/4$ & $2q(N-q+1)$\\
\hline
\end{tabular}
\caption{The spectrum of easy bosons where $\ell=1,2,\dots,q-3/2$. $Y^{\tilde{\alpha}}(\searrow)$ and $Y^{\dagger}_{\tilde{\alpha}}(\searrow)$ are massless. The multiplicity refers to the number of real fields and includes both $\tilde{\alpha}=3$ and
$\tilde{\alpha}=4$.   \label{tab spec EB}}\quad\quad
\end{table}

  \subsection{The easy fermions}
        
    The Gaussian action of easy fermions reads
    \begin{equation}
    	\begin{aligned}
    		\mathcal{L}_{\text{f,e}}=\text{tr}\bigg\{ i\bar{\psi}^{\alpha\dagger}\gamma^{\mu}\partial_{\mu}\psi_{\alpha} &-\frac{1}{2z}y_{\alpha}^{\dagger}y^{\alpha}\bar{\psi}^{\beta\dagger}\psi_{\beta}+\frac{1}{2z}\bar{\psi}^{\alpha\dagger}y^{\beta} y_{\beta}^{\dagger}\psi_{\alpha}\\ &+\frac{1}{z}y_{\alpha}^{\dagger}y^{\beta}\bar{\psi}^{\alpha\dagger}\psi_{\beta}-\frac{1}{z}\bar{\psi}^{\alpha\dagger} y^{\beta} y_{\alpha}^{\dagger} \psi_{\beta}\bigg\}.
    	\end{aligned}
    \end{equation}
    Expressing the bilinear form of the classical fields in terms of $\mathfrak{su}(2)$ representations as shown in Sec. \ref{BPS equation}, one can write the mass terms above as
    \begin{equation}
    \begin{aligned}
        &\frac{1}{z}\text{tr}\left( -\frac{1}{2}y_{\alpha}^{\dagger}y^{\alpha}\bar{\psi}^{\beta\dagger}\psi_{\beta}+\frac{1}{2}\bar{\psi}^{\alpha\dagger}y^{\beta} y_{\beta}^{\dagger}\psi_{\alpha}+y_{\alpha}^{\dagger}y^{\beta}\bar{\psi}^{\alpha\dagger}\psi_{\beta}-\bar{\psi}^{\alpha\dagger} y^{\beta} y_{\alpha}^{\dagger} \psi_{\beta} \right)=\frac{1}{z}\text{tr} \left( \bar{\psi}^{\alpha\dagger}(\sigma_i)_{\alpha}^{\ \beta}\Tilde{L}_i \psi_{\beta} \right),
        \end{aligned}
    \end{equation}
    where we used eq.\ (\ref{2.16}) and defined
    \begin{equation}
        \Tilde{L}_i X=-t_i^T X-X \hat{t}_i,
    \end{equation}
    As $-t_i^T$ ($-\hat{t}_i$ hitting from the right) furnishes the $\mathfrak{su}(2)$ representation equivalent to $\pi_{q-1}$ ($\pi_q$), one finds $\tilde{L}_i$ is again in the $\pi_{q-1}\otimes\pi_{q}$ representation.  By block decomposition we can further write
    \begin{equation}
\begin{aligned}
    \frac{1}{z}\text{tr} \left( \bar{\psi}^{\alpha\dagger}(\sigma_i)_{\alpha}^{\ \beta}\Tilde{L}_i \psi_{\beta} \right)=&\frac{1}{z}\text{tr} \left( \bar{\psi}^{\alpha\dagger}(\nwarrow)(\sigma_i)_{\alpha}^{\ \beta}\Tilde{L}_i \psi_{\beta}(\nwarrow) \right)
    \\
    +&\frac{1}{z} \text{tr}\left( \bar{\psi}^{\alpha\dagger}(\swarrow)(\sigma_i)_{\alpha}^{\ \beta} (-t_i^T) \psi_{\beta}(\nearrow) \right)
    \\
    +&\frac{1}{z} \text{tr}\left( \bar{\psi}^{\alpha\dagger}(\nearrow)\psi_{\beta}(\swarrow)(\sigma_i)_{\alpha}^{\ \beta}(-\hat{t}_i)  \right).
    \end{aligned}
\end{equation}
    Now we conclude that 
    \begin{itemize}
		\item  $\bar{\psi}^{\alpha\dagger}(\nwarrow)$ and $\psi_{\alpha}(\nwarrow)$ are mixed in the $\pi_2\otimes\pi_{q-1}\otimes\pi_{q}$ representation specified by  $-2\pi_2(s_i)\otimes(\pi_{q-1}\otimes\pi_q)(s_i)$,
		\item  $\bar{\psi}^{\alpha\dagger}(\swarrow)$ and $\psi_{\alpha}(\nearrow)$ are mixed in the $\pi_2\otimes\pi_{q-1}$ representation specified by 
		 $-2\pi_2(s_i)\otimes\pi_{q-1}(s_i) $,
		\item $\bar{\psi}^{\alpha\dagger}(\nearrow)$ and $\psi_{\alpha}(\swarrow)$ are mixed in the $\pi_2\otimes\pi_{q}$ representation specified by
		$  -2\pi_2(s_i)\otimes\pi_q(s_i) $,
	\end{itemize}
    where $s_i$ denote the corresponding $\mathfrak{su}(2)$ generators. The factor of 2 appears in the mixing operator since the 2-dimensional irreducible representation is furnished by $\sigma_i/2$, and the minus sign arises from the definition of the fermionic mass term.

    Hence, the mixing of easy fermions can again be resolved by $\mathfrak{su}(2)$ representation theory, which yields
    \begin{equation}
        \pi_2(s_i)\otimes\pi_d(s_i)=\frac{d-1}{4}\mathbbm{1}_{d+1}\,\oplus \,\frac{-d-1}{4}\mathbbm{1}_{d-1}.
    \end{equation}
    Then by decomposing $\pi_{q-1}\otimes\pi_q$ one further finds
    \begin{equation}
        \pi_2(s_i)\otimes(\pi_{q-1}\otimes\pi_q)(s_i)=\bigoplus_{r=1}^{q-1} \frac{2r-1}{4}\mathbbm{1}_{2r+1}\oplus\frac{-2r-1}{4}\mathbbm{1}_{2r-1}.
    \end{equation}
    With the above formulas, it is straightforward to write down the spectrum of all easy fermions 
    and the result  is shown in Table \ref{tab spec EF}.
    \begin{table}[htbp]
\centering
\begin{tabular}{c|c}
\multicolumn{2}{c}{$\psi_{\alpha}(\nwarrow)$} 
\\
\hline
mass&multiplicity\\
\hline
$-r+1/2$ & $2r+1$\\
\hline
$r+1/2$ & $2r-1$\\
\hline
\end{tabular}\\[10pt] 
\begin{tabular}{c|c}
\multicolumn{2}{c}{$\psi_{\alpha}(\nearrow)$} 
\\
\hline
mass&multiplicity\\
\hline
$-(q-2)/2$ & $q(N-q)$\\
\hline
$q/2$ & $(q-2)(N-q)$\\
\hline
\end{tabular}\\[10pt]
\begin{tabular}{c|c}
\multicolumn{2}{c}{$\psi_{\alpha}(\swarrow)$} 
\\
\hline
mass&multiplicity\\
\hline
$-(q-1)/2$ & $(q+1)(N-q+1)$\\
\hline
$(q+1)/2$ & $(q-1)(N-q+1)$\\
\hline
\end{tabular}
\caption{The spectrum of easy fermions where $r=1,2,\dots,q-1$. $\psi_{\alpha}(\searrow)$ and $\bar{\psi}^{\alpha\dagger}(\searrow)$ are massless.  \label{tab spec EF}}\quad\quad
\end{table}

Using the parameter $\nu$ to parametrize the spectrum of the scalars
we can organize the spectrum of easy fields as in table~\ref{tab easy}.
 We see that the spectrum of the easy fields indeed carry the characteristics of a supersymmetric spectrum with the
 conformal dimensions of bosons and fermions differing by $\frac{1}{2}$.       

\begin{table}
        \centering
    		\begin{tabular}{|c|c|c|}
    			\hline
    			Multiplicity& $\nu$ of $Y^{\tilde{\alpha}}$ & $m$ of $\psi_{\alpha}$  \\ \hline
    			$2r+1$ & $r$ & $-r+\frac{1}{2}$ \\[3pt]
    			$2r-1$ & $r$ & $r+\frac{1}{2}$ \\[3pt] 
    			$(q+1)(N-q+1)$ & $\frac{q}{2}$ & $-\frac{q-1}{2}$ \\[3pt] 
    			$(q-1)(N-q+1)$ & $\frac{q}{2}$ & $\frac{q+1}{2}$ \\[3pt] 
    			$q(N-q)$ & $\frac{q-1}{2}$ & $-\frac{q-2}{2}$ \\[3pt] 
    			$(q-2)(N-q)$ & $\frac{q-1}{2}$ & $\frac{q}{2}$ \\[3pt] 
    			$2(N-q)(N-q+1)$ & $\frac{1}{2}$ & 0 \\ \hline
    		\end{tabular}
\caption{The spectrum of easy fields witht $r=1,\ldots, q-1$. \label{tab easy}}
 \end{table}

\section{Diagonalization of the complicated fields  \label{complicated field}}
       
        In this section, we will proceed with the diagonalization of the complicated fields. For the complicated fermions the mixing 
        appears only in the mass terms and we can proceed as in the previous section. In the case of the complicated bosons there is mixing both in the mass terms and the kinetic terms due to the Chern Simons interaction which makes this case
more difficult to handle. We start with the easier case.
        
    \subsection{The complicated fermions}

        The Gaussian action for the complicated fermions reads
    \begin{equation}
    	\begin{aligned}
    		\mathcal{L}_{\text{f,c}}=\text{tr}\bigg\{ i\bar{\psi}^{\Tilde{\alpha}\dagger}\gamma^{\mu}\partial_{\mu}\psi_{\Tilde{\alpha}}& -\frac{1}{2z}y_{\alpha}^{\dagger}y^{\alpha}\bar{\psi}^{\Tilde{\alpha}\dagger}\psi_{\Tilde{\alpha}}+\frac{1}{2z}\bar{\psi}^{\Tilde{\alpha}\dagger}y^{\alpha} y_{\alpha}^{\dagger}\psi_{\Tilde{\alpha}}\\
            &+\frac{i}{2z}\epsilon^{\alpha \Tilde{\alpha} \beta \Tilde{\beta}}y_{\alpha}^{\dagger}\bar{\psi}_{\Tilde{\alpha}} y_{\beta}^{\dagger} \psi_{\Tilde{\beta}}-\frac{i}{2z}\epsilon_{\alpha \Tilde{\alpha} \beta \Tilde{\beta}}y^{\alpha}\bar{\psi}^{\Tilde{\alpha}\dagger} y^{\beta} \psi^{\Tilde{\beta}\dagger}\bigg\}.
    	\end{aligned}
    \end{equation}
    The mixing takes place only in the mass terms.  After  block decomposition of the fields one finds the following terms involving the off-diagonal blocks
    \begin{equation}
        \begin{aligned}
            &\frac{1}{z}\text{tr}\left\{-\frac{q-1}{2}\bar{\psi}^{\Tilde{\alpha}\dagger}(\nearrow) \psi_{\Tilde{\alpha}}(\swarrow)+\frac{q}{2}\bar{\psi}^{\Tilde{\alpha}\dagger}(\swarrow) \psi_{\Tilde{\alpha}}(\nearrow) \right\},
        \end{aligned}
    \end{equation}
    which are already diagonal. Hence, we need only discuss the diagonalization of the $(\nwarrow)$-components of the field, which are mixed in the following way
    \begin{equation}
        \begin{aligned}
            &\frac{1}{z}\bigg\{ \frac{1}{2}\bar{\psi}^{\Tilde{\alpha}\dagger}(\nwarrow)\psi_{\Tilde{\alpha}}(\nwarrow)+\frac{i}{2}\epsilon^{\alpha \Tilde{\alpha} \beta \Tilde{\beta}}y_{\alpha}^{\dagger}\bar{\psi}_{\Tilde{\alpha}}(\nwarrow)y_{\beta}^{\dagger} \psi_{\Tilde{\beta}}(\nwarrow)-\frac{i}{2}\epsilon_{\alpha \Tilde{\alpha} \beta \Tilde{\beta}}y^{\alpha}\bar{\psi}^{\Tilde{\alpha}\dagger}(\nwarrow) y^{\beta} \psi^{\Tilde{\beta}\dagger}(\nwarrow)\bigg\}\\
            =&\frac{1}{z}\bigg\{ \frac{1}{2}\bar{\psi}^{\Tilde{\alpha}\dagger}(\nwarrow)\psi_{\Tilde{\alpha}}(\nwarrow)-\frac{1}{2}\left( \bar{\psi}^{\Tilde{\alpha}\dagger*}(\nwarrow)\hat{F}\psi_{\Tilde{\alpha}}(\nwarrow)+\bar{\psi}^{\Tilde{\alpha}\dagger}(\nwarrow)\hat{F}\psi_{\Tilde{\alpha}}^{*}(\nwarrow)\right)\bigg\},
        \end{aligned}
    \end{equation}
    where we used the fact that the upper and lower flavour indices are related by $\epsilon^{\Tilde{\alpha}\Tilde{\beta}}$ since they transform under SU(2). Precisely, we use the convention 
    \begin{equation}
        \psi^{\Tilde{\alpha}}=\epsilon^{\Tilde{\alpha}\Tilde{\beta}}\psi_{\Tilde{\beta
        }}, \quad \epsilon^{\Tilde{\alpha}\Tilde{\beta}}=\begin{pmatrix}
            0 & -1 \\ 1 & 0
        \end{pmatrix}\quad \Rightarrow \quad
        \psi^3=-\psi_4, \quad \psi^4=\psi_3.
    \end{equation} 
    The operator $\hat{F}$, which specifies the mixing in the gauge space, is defined by
    \begin{equation}
        \hat{F} \phi:=iy^1\phi^{\dagger*} y^2-iy^2\phi^{\dagger*}y^1.
    \end{equation}
    Also, it can be verified that $\hat{F}$ commutes with $L_2$ defined in eq. (\ref{def of Li}) as well as with the Casimir $L_i L_i$. A numerical investigation is compatible with the following result
    \begin{equation}\label{numerical}
        \hat{F} R_{j+1/2}^{n+1/2} = (-1)^{j-n}(j+1) R_{j+1/2}^{n+1/2},
    \end{equation}
    where the $R$ is the modified fuzzy spherical harmonics for $L_2$ with $j=0,1,\dots,q-2$, i.e. the analog of the $L_3$ eigenstate $T$ introduced in eq.\ (\ref{L3 on T}). 

    Let us proceed with disentangling the mixing of the $(\nwarrow)$ block. First one can diagonalize $\hat{F}$ by transforming to the $R$-basis, where the fermionic fields are expanded as
    \begin{equation}
        \psi_{\tilde{\alpha}}=\psi_{\tilde{\alpha},jn}R_{j+1/2}^{n+1/2}.
    \end{equation}
    Then for a given $(j,n)$ one finds
    \begin{equation}
    \begin{aligned}
        \frac{1}{z}\left(\frac{1}{2}\bar{\psi}^{\Tilde{\alpha}*}_{jn}\psi_{\Tilde{\alpha},jn}+\frac{i}{2}F_{jn}\left( \bar{\psi}_{jn}^{\Tilde{\alpha}}\psi_{\Tilde{\alpha},jn}+\bar{\psi}_{jn}^{\Tilde{\alpha}*}\psi_{jn}^* \right)\right),
    \end{aligned}
    \end{equation}
    where we denote the $\hat{F}$ eigenvalue as $F_{jn}=(-1)^{j-n}(j+1)$. Suppressing all $jn$'s on field components as well as flavour indices, one finds a concise expression where $\psi$ is a usual fermion without any gauge or flavour d.o.f.\footnote{We are using the convention $\bar{\psi}=\psi^T\gamma^0$ and $\gamma^0=\sigma^3$. Hence, $\bar{\psi}^*$ is the usual Dirac adjoint of $\psi$.}
    \begin{equation}
    \frac{1}{z}\left(\frac{1}{2}\bar{\psi}^*\psi-\frac{1}{2}F_{jn}\left( \bar{\psi}\psi+\bar{\psi}^*\psi^{*} \right)\right).
    \end{equation}
    Therefore, we are led to separating the real and imaginary parts of the fermion, i.e.\ we define $\psi_R$ and $\psi_I$ by
    \begin{equation}
        \psi=\psi_R+i\psi_I\Rightarrow \bar{\psi}^*=\bar{\psi}_R^*-i\bar{\psi}_I^*.
    \end{equation}
    Note that $\psi_{R,I}$ is real so $\bar{\psi}_{R,I}^*=\psi_{R,I}^T \gamma^0=\bar{\psi}_{R,I}$. In terms of ${\psi}_{R,I}$, the mixing reads
    \begin{equation}
        \begin{aligned}
            &(\frac{1}{2}-F_{jn})\bar{\psi}_R^*\psi_R+(\frac{1}{2}+F_{jn})\bar{\psi}_I^*\psi_I,\\
        \end{aligned}
    \end{equation}
    where it is straightforward to read off the masses $-(\frac{1}{2}\pm F_{jn})$. To compare that with the bosonic masses, we let $\ell=j+1$, and thus for a certain $\ell$ (and $\tilde{\alpha}$) one finds the mass $-\ell-1/2$ with the multiplicity $2\ell$ and the mass $\ell-1/2$ with the same multiplicity. 
    
    Based on the above discussion, we arrive at  table~\ref{tab spec CF} for the spectrum of the complicated fermions. It would
  be interesting to complete the analysis with an analytical derivation of eq.~(\ref{numerical}).

    \begin{table}[htbp]
\centering
\begin{tabular}{c|c}
\multicolumn{2}{c}{$\psi_{\tilde{\alpha}}(\nwarrow)-\bar{\psi}^{\tilde{\alpha}\dagger}(\nwarrow)$}\\
\hline
mass&multiplicity\\
\hline
$\ell-1/2$ & $4\ell$\\
\hline
$-\ell-1/2$ & $4\ell$\\
\hline
\end{tabular}\\[10pt] 
\begin{tabular}{c|c}
\multicolumn{2}{c}{$\psi_{\tilde{{\alpha}}}(\nearrow)$}\\
\hline
mass&multiplicity\\
\hline
$-q/2$ & $2(q-1)(N-q)$\\
\hline
\end{tabular}\\[10pt]
\begin{tabular}{c|c}
\multicolumn{2}{c}{$\psi_{\tilde{\alpha}}(\swarrow)$}\\
\hline
mass&multiplicity\\
\hline
$(q-1)/2$ & $2q(N-q+1)$\\
\hline
\end{tabular}
\caption{The spectrum of complicated fermions where $\ell=1,2,\dots,q-1$ and $\psi_{\tilde{\alpha}}(\searrow)$ are massless. The minus sign comes from the definition of the fermionic mass term. The multiplicity counts $\tilde{\alpha}=3,4$ for each block, and counts the real and complex parts for the diagonal block since they are separated out in the diagonalization. 
\label{tab spec CF}}\quad\quad
\end{table}

\subsection{The complicated bosons}
Next, we turn to the complicated bosons that have Gaussian action
\begin{equation}\label{Lagrangian of CB}
		\begin{aligned}
			S_{\text{b,c}}=\frac{k}{4\pi}\int d^3x\,\text{tr}\bigg \{ &-\epsilon^{\mu\rho\nu}A_{\mu}\partial_{\rho}A_{\nu} + \epsilon^{\mu\rho\nu}\hat{A}_{\mu}\partial_{\rho}\hat{A}_{\nu}\\
   &-\frac{2}{z}\hat{A}_{\mu}y_{\alpha}^{\dagger}A^{\mu}y^{\alpha}+\frac{1}{z}y_{\alpha}^{\dagger}A_{\mu}A^{\mu}y^{\alpha}+\frac{1}{z}\hat{A}_{\mu}y_{\alpha}^{\dagger}y^{\alpha}\hat{A}^{\mu}\\
			&+\left( \partial_{\mu}Y^{\alpha} \right)\partial^{\mu}Y_{\alpha}^{\dagger}+\left( Y^{\dagger}-Y\ \text{mixing} \right)\\
			&+\frac{i}{z^{3/2}}y^{\alpha}\left(Y_{\alpha}^{\dagger}A^z- \hat{A}^z Y_{\alpha}^{\dagger}\right)+\frac{i}{z^{3/2}}y_{\alpha}^{\dagger}\left(Y^{\alpha}\hat{A}^z- A^z Y^{\alpha} \right)\\
			&+\frac{i}{z^{1/2}}(\partial_{\mu}A^{\mu})\left( Y^{\alpha}y_{\alpha}^{\dagger}-y^{\alpha}Y_{\alpha}^{\dagger} \right)+\frac{i}{z^{1/2}}(\partial_{\mu}\hat{A}^{\mu})\left( Y_{\alpha}^{\dagger}y^{\alpha}-y_{\alpha}^{\dagger}Y^{\alpha} \right) \bigg\}.
		\end{aligned}
	\end{equation}
We note that in this case the mixing terms involve kinetic terms, both in the form of Chern Simons terms and terms originating from the covariant derivative of the scalar fields. The latter terms could be elminated by working in 
background field gauge as in~\cite{Buhl-Mortensen:2016jqo} but we find that the gauge fixing does not simplify the mixing
problem. In order to find the spectrum of quantum fluctuations of the fields involved we will solve the relevant field equations
of motion and later handle the gauge fixing by imposing convenient gauge conditions on the solutions.  We can read off the spectrum from the scaling behavior of the solutions of the equations of motion near the defect as explained in section~\ref{Exp-spectrum}.
The solutions are expected to be given in terms of modified Bessel functions depending on the transverse coordinate $z$, and the  parameter $\nu$ is easily identified as the index of the relevant Bessel function. Finding the solutions of the equations of motion also
makes it possible later to construct the propagators by pasting together solutions with the appropriate boundary behaviour, cf.\ eq.\ (\ref{eq:PropagatorBessel}).

Again, the diagonal blocks ($\nwarrow, \searrow$) and the off-diagonal blocks ($\nearrow, \swarrow$) do not couple to each other. The mixing problem is much simpler for the off-diagonal blocks since terms where quantum fields are not adjacent 
do not contribute in this case. In particular,  $m_{YY}$ in the $Y^{\dagger}\mbox{-}Y$ mixing can be ignored (cf.\ eqs.\ 
(\ref{YAYA})-(\ref{YY})). Furthermore, as we will see below, $Y^{\alpha}(\swarrow)$ mixes only with $A^{\mu}(\swarrow)$ while $Y^{\alpha}(\nearrow)$ mixes only with $\hat{A}^{\mu}(\nearrow)$. In the following, we first consider the off-diagonal blocks and subsequently the upper diagonal blocks, (${\nwarrow})$, while the $(\searrow)$-blocks are easily seen to be massless.

\subsubsection{$Y(\swarrow)-A(\swarrow)$ mixing}

One finds that $Y(\swarrow),A(\swarrow)$ (and so $Y(\nearrow), A(\nearrow)$) form  closed subsets  of fields. The former
two fields are coupled via the Lagrangian
\begin{equation}\label{Lagrangian of Ysw}
    	\begin{aligned}
    		\mathcal{L} = \text{tr} \biggr[ &Y^{\alpha}(\swarrow)\left( -\partial_{\mu}\partial^{\mu}\delta_{\alpha}^{\beta} -\frac{(q-1)(q-2)}{8z^2}\delta_{\alpha}^{\beta} -\frac{q-4}{2z^2}\hat{t}_i\frac{1}{2}(\sigma_i)_{\alpha}^{\ \beta}\ \right)Y_{\beta}^{\dagger}(\nearrow)\\
    		&+A^{\mu}(\swarrow)\left(-2\epsilon_{\mu\rho\nu}\partial^{\rho}+\frac{q}{z}\eta_{\mu\nu}\right)A^{\nu}(\nearrow)\\
    		&-\frac{i}{z^{3/2}}Y^{\alpha}(\swarrow)y_{\alpha}^{\dagger}A^z(\nearrow)+\frac{i}{z^{3/2}}A^z(\swarrow)y^{\alpha}Y_{\alpha}^{\dagger}(\nearrow)\\
      &+\frac{i}{z^{1/2}}Y^{\alpha}(\swarrow)y_{\alpha}^{\dagger}\left(\partial_{\mu}A^{\mu}(\nearrow)\right)-\frac{i}{z^{1/2}}\left(\partial_{\mu}A^{\mu}(\swarrow)\right)y^{\alpha}Y_{\alpha}^{\dagger}(\nearrow) \biggr],
    	\end{aligned}
    \end{equation}
    where we have replaced every bilinear combination of classical fields with the solutions to the BPS equations, shown in Section \ref{BPS equation}, and used the $\mathfrak{su}(2)$ algebra. From eq.\ (\ref{Lagrangian of Ysw}) one finds the equations of motion 
    \begin{equation}\label{EOM of Y}
    \begin{aligned}
    \left[\partial_{\mu}\partial^{\mu}\delta_{\alpha}^{\beta}+\frac{(q-1)(q-2)}{8z^2}\delta_{\alpha}^{\beta}+\frac{q-4}{2z^2}\hat{t}_i\frac{1}{2}(\sigma_i)_{\alpha}^{\ \beta}\right]Y_{\beta}^{\dagger}(\nearrow)+\left(\frac{i}{z^{3/2}}\delta^{z}_{\mu}-\frac{i}{z^{1/2}}\partial_{\mu}\right)y_{\alpha}^{\dagger}A^{\mu}(\nearrow)=0,
    \end{aligned}
    \end{equation}
    \begin{equation}\label{EOMA}
    	\left(2\epsilon_{\mu\rho\nu}\partial^{\rho}-\frac{q}{z}\eta_{\mu\nu}\right)A^{\nu}(\nearrow)-\left(\frac{i}{z^{1/2}}\partial_{\mu}+\frac{i}{2z^{3/2}}\delta_{\mu z}\right)y^{\alpha}Y_{\alpha}^{\dagger}(\nearrow)=0.
    \end{equation}
  Now we first solve the EOM of the gauge fields, eq.\ (\ref{EOMA}), and then plug this solution into the EOM of
      scalars, eq.\ (\ref{EOM of Y}),
      to find the complete solution.\footnote{One can multiply $y^{\alpha}$ to both sides of the EOM of the scalar field and get
    \begin{equation*}
    	\left( \partial^{\mu}\partial_{\mu}+\frac{3}{4z^2} \right)y^{\alpha}Y_{\alpha}^{\dagger}(\nearrow)+iq\left( \frac{1}{z^{3/2}}\delta^{z}_{\mu}-\frac{1}{z^{1/2}}\partial_{\mu} \right)A^{\mu}(\nearrow)=0.
    \end{equation*}
    Then multiplying $\partial^{\mu}$ to both sides of the EOM of the gauge field yields the same equation, which shows the consistency of the set of equations.}   From the generic results in Appendix \ref{AppendixCS}, one obtains the following solution to the gauge field EOM
   \begin{equation}
   	\begin{aligned}
   		&A^a(\nearrow)=-\frac{i}{2ik}\epsilon^{ab}k_b f K_{\frac{q}{2}}(ikz),\\
   		&A^z(\nearrow)=\frac{1}{4}f\left[ K_{\frac{q}{2}-1}(ikz)-K_{\frac{q}{2}+1}(ikz) \right],\\
   		&y^{\alpha}Y_{\alpha}^{\dagger}(\nearrow)=\frac{\sqrt{z}}{2i}f\left[ K_{\frac{q}{2}-1}(ikz)+K_{\frac{q}{2}+1}(ikz) \right],
   	\end{aligned}
   \end{equation}
   where it is understood that we are working in momentum space for the longitudinal directions, $a=1,2$, with momentum 
   variables $k^0,k^1$ and $ik=\sqrt{-k_a k^a}$.
   The $(q-1)\times(N-q+1)$ matrix $f$ should be determined by normalization conditions. The arguments of our modified Bessel functions
   reflect the fact that we are working in Lorentzian signature.

   Next, we  solve the scalar EOM using the result above. By a rescaling $Y_{\alpha}^{\dagger}(\nearrow)=\sqrt{z}W_{\alpha}^{\dagger}(\nearrow)$, one can write the EOM of the $Y$-fields as
   \begin{equation}\label{hatB action}
   	 \left[ -\hat{B}\delta_{\alpha}^{\beta}+\frac{1}{4}\delta_{\alpha}^{\beta}+\frac{(q-1)(q-2)}{8}\delta_{\alpha}^{\beta}+\frac{q-4}{2}\hat{t}_i\frac{1}{2}(\sigma_i)_{\alpha}^{\ \beta} \right]W_{\beta}^{\dagger}(\nearrow)+iy_{\alpha}^{\dagger}(1-z\partial_z)A^z(\nearrow)=0,
   \end{equation}
   where for convenience we define $\hat{B}=z^2\partial_z^2+z\partial_z-z^2\partial_a\partial^a$, the derivative part of the Bessel equation that has the Bessel function as the eigenfunction
   \begin{equation}
       \hat{B}\,e^{-ik_ax^a}K_{\nu}(ikz)=\nu^2 e^{-ik_ax^a}K_{\nu}(ikz).
   \end{equation}
   The second term of the EOM can be computed by the recurrence relation of the modified Bessel function eq.\ (\ref{C3}) and eq.\ (\ref{C4}). For the first term, one can perform a transformation on the gauge-flavor space to diagonalize $\hat{t}_i\sigma_i/2$, which is the decomposition of the reducible representation $\pi_{2}(s_i)\otimes\pi_{q}(s_i)\cong\pi_{q+1}\oplus\pi_{q-1}$ as before. Then the mixing in the gauge-flavor space is transformed to
   \begin{equation}
   	\begin{aligned}
   		&\frac{1}{4}\delta_{\alpha}^{\beta}+\frac{(q-1)(q-2)}{8}\delta_{\alpha}^{\beta}+\frac{q-4}{2}\hat{t}_i\frac{1}{2}(\sigma_i)_{\alpha}^{\ \beta}\rightarrow \left( \frac{q}{2}-1 \right)^2\mathbbm{1}_{q+1}\, \oplus \, \mathbbm{1}_{q-1}.
   	\end{aligned}
   \end{equation}
   We also use the following notation to denote the transformed fields
   \begin{equation}
   	W_{\alpha}^{\dagger}(\nearrow)\rightarrow W_{[q+1]}^{\dagger}(\nearrow)\oplus W_{[q-1]}^{\dagger}(\nearrow); \quad y_{\alpha}^{\dagger}f\rightarrow y^{\dagger}_{[q+1]}f\oplus y^{\dagger}_{[q-1]}f.
   \end{equation}
   We notice that  $W_{\alpha}^{\dagger}(\nearrow)$ is a $q\times(N-q+1)$ matrix with 2-dimensional flavor DOF, and the transformation acts on the $(2\times q)$-dimensional tensor product space of the row index and the flavor index with the column index unchanged. Hence, $W_{[q+1]}^{\dagger}(\nearrow)$ is a $(q+1)\times(N-q+1)$ matrix in the $\pi_{q+1}$ irreducible representation. The same idea is applied to the other matrix fields. One also finds that $f$ stays invariant under the transformation since only the row and flavor indices of $y_{\alpha}^{\dagger}$ should be transformed. This decomposition decouples the fields in different irreducible representations, so we discuss them separately in what follows.
   
   The fields in the $\pi_{q-1}$ representation have the EOM
   \begin{equation}
   	(-\hat{B}+1)W_{[q-1]}^{\dagger}(\nearrow)+\frac{i}{4}y_{[q-1]}^{\dagger}f\left[ (2-\frac{q}{2})K_{\frac{q}{2}-1}(ikz)-(2+\frac{q}{2})K_{\frac{q}{2}+1}(ikz) \right]=0,
   \end{equation}
 and  it is not hard to obtain the solution
   \begin{equation}
   	W_{[q-1]}^{\dagger}(\nearrow)=\frac{1}{2i}\frac{1}{q}y^{\dagger}_{[q-1]}f\left[ K_{\frac{q}{2}-1}(ikz)+K_{\frac{q}{2}+1}(ikz) \right].
    \end{equation}
The fields in the $\pi_{q+1}$ representation have the EOM
    \begin{equation}
    	\left[-\hat{B}+(\frac{q}{2}-1)^2\right]W_{[q+1]}^{\dagger}(\nearrow)+\frac{i}{4}y_{[q+1]}^{\dagger}f\left[ (2-\frac{q}{2})K_{\frac{q}{2}-1}(ikz)-(2+\frac{q}{2})K_{\frac{q}{2}+1}(ikz) \right]=0.
    \end{equation}
    The $K_{\frac{q}{2}-1}(ikz)$ in the second term implies there should also be such Bessel function in $W_{[q+1]}^{\dagger}(\nearrow)$ since $\hat{B}$ does not change the order of the Bessel function. However, one by eq. (\ref{hatB action}) finds the square bracket in front of $W_{[q+1]}^{\dagger}(\nearrow)$ annihilates $K_{\frac{q}{2}-1}(ikz)$. Therefore, the only possible solution is
    \begin{equation}
    \label{4.21}
        W_{[q+1]}^{\dagger}(\nearrow)=h_{[q+1]}K_{\frac{q}{2}-1}(ikz), \quad y^{\dagger}_{[q+1]}f=0,
    \end{equation}
    where $h_{[q+1]}$ is a $(q+1)\times(N-q+1)$ matrix independent of $f$ and should also be determined by normalization conditions. The second equation looks like an unexpected restriction on $f$, but it is actually a natural consequence of the fact that the decomposition of $\pi_2\otimes \pi_q\cong\pi_{q+1}\oplus\pi_{q-1}$, while $y_{\alpha}^{\dagger}$ is the eigenvector of the operator
    \begin{equation}
        \hat{t}_i\frac{1}{2}(\sigma_i)_{\alpha}^{\ \beta}y_{\beta}^{\dagger}=-\frac{q+1}{4}y_{\alpha}^{\dagger}.
    \end{equation}
Thus $y_{\alpha}^{\dagger}$ itself is in the $(q-1)$-dimensional irrp, which means $y^{\dagger}_{[q+1]}=0$.

    We can reorganize the solutions in terms of a single Bessel function
   \begin{equation}
   	\begin{aligned}
   	&A^a(\nearrow)=-\frac{i}{2ik}\epsilon^{ab}k_b f K_{\frac{q}{2}}(ikz),\\
   	&X^+(\nearrow) = f K_{\frac{q}{2}-1}(ikz),\\
   	&X^-(\nearrow) =-fK_{\frac{q}{2}+1}(ikz),\\
   	&Y_{[q+1]}^{\dagger}(\nearrow)=\sqrt{z}h_{[q+1]}K_{\frac{q}{2}-1}(ikz),\\
   	\end{aligned}
   \end{equation}
   where
   \begin{equation}
   	X^{\pm}(\nearrow)=2A^z(\nearrow)\pm \frac{i}{\sqrt{z}}y^{\alpha}Y_{\alpha}^{\dagger}(\nearrow).
   \end{equation}
   As explained in section~\ref{Exp-spectrum} we can read off the value of the parameter $\nu$ from the 
   index of the modified Bessel functions.   The results are shown in Table \ref{tab spec CB Ysw}.
\begin{table}[htbp]
\centering
\begin{tabular}{c|c}
\multicolumn{2}{c}{$Y_{\alpha}^{\dagger}(\nearrow)-A^{\mu}(\nearrow)$,\ $Y^{\alpha}(\swarrow)-A^{\mu}(\swarrow)$}\\
\hline
$\nu$ &multiplicity\\
\hline
$q/2$ & $2(q-1)(N-q+1)$\\
\hline
$q/2-1$ & $2q(N-q+1)$\\
\hline
$q/2+1$ & $(q-1)(N-q+1)$ \\
\hline
\end{tabular}\\[10pt] 
\caption{The spectrum of the off-diagonal blocks $Y_{\alpha}^{\dagger}(\nearrow), A^{\mu}(\nearrow)$ and $Y^{\alpha}(\swarrow)$, $A^{\mu}(\swarrow)$. Notice that the total multiplicity adds up to $(10\,q-6)(N-q+1)$ which exactly matches the number of independent
real field components in the fields involved. 
\label{tab spec CB Ysw}}\quad\quad
\end{table}

   \subsubsection{$Y(\nearrow)-\hat{A}(\nearrow)$ mixing}

   In this section, we discuss the mixing between $Y(\nearrow)$ and $\hat{A}(\nearrow)$, and the idea is the same as that for the off-diagonal blocks discussed in the previous subsection. The decoupled Lagrangian reads
   \begin{equation}
   	\begin{aligned}
   		\mathcal{L} = \text{tr} \biggr[ &Y_{\alpha}^{\dagger}(\swarrow)\left( -\partial_{\mu}\partial^{\mu}\delta^{\alpha}_{\beta} -\frac{q(q+1)}{8z^2}\delta^{\alpha}_{\beta} +\frac{q+3}{2z^2}t_i\frac{1}{2}(\sigma_i)^{\alpha}_{\ \beta}\ \right)Y^{\beta}(\nearrow)\\
   		&+\hat{A}^{\mu}(\swarrow)\left(2\epsilon_{\mu\rho\nu}\partial^{\rho}+\frac{q-1}{z}\eta_{\mu\nu}\right)\hat{A}^{\nu}(\nearrow)\\
   		&-\frac{i}{z^{3/2}}Y_{\alpha}^{\dagger}(\swarrow)y^{\alpha}\hat{A}^z(\nearrow)+\frac{i}{z^{3/2}}\hat{A}^z(\swarrow)y_{\alpha}^{\dagger}Y^{\alpha}(\nearrow)\\
     &+\frac{i}{z^{1/2}}Y_{\alpha}^{\dagger}(\swarrow)y^{\alpha}\left( \partial_{\mu}\hat{A}^{\mu}(\nearrow)\right)-\frac{i}{z^{1/2}}\left( \partial_{\mu}\hat{A}^{\mu}(\swarrow) \right)y_{\alpha}^{\dagger}Y^{\alpha}(\nearrow) \biggr].
   	\end{aligned}
   \end{equation}
   And one finds the EOM
   \begin{equation}
   	\left[ \partial_{\mu}\partial^{\mu}\delta^{\alpha}_{\ \beta}+\frac{q(q+1)}{8z^2}\delta^{\alpha}_{\ \beta}-\frac{q+3}{2z^2}t_i\frac{1}{2}(\sigma_i)^{\alpha}_{\ \beta} \right]Y^{\beta}(\nearrow)+iy^{\alpha}\left( \frac{1}{z^{3/2}}\delta^{z}_{\mu}-\frac{1}{z^{1/2}}\partial_{\mu} \right)\hat{A}^{\mu}(\nearrow)=0,
   \end{equation}
   \begin{equation}
   	\left(2\epsilon_{\mu\rho\nu}\partial^{\rho}+\frac{q-1}{z}\eta_{\mu\nu}\right)\hat{A}^{\nu}(\nearrow)+i\left( \frac{1}{z^{1/2}}\partial_{\mu}+\frac{1}{2z^{3/2}}\delta^{z}_{\mu} \right)y_{\alpha}^{\dagger}Y^{\alpha}(\nearrow)=0.
   \end{equation}
   The gauge field EOM again has the form of that in eq.\ (\ref{massive CS EOM}), hence we have the solution
   \begin{equation}
   	\begin{aligned}
   		&\hat{A}^a=\frac{i}{2ik}\epsilon^{ab}k_b\hat{f} K_{\frac{q-1}{2}}(ikz),\\
   		&\hat{A}^z=\frac{1}{4}\hat{f}\left[ K_{\frac{q-1}{2}-1}(ikz)-K_{\frac{q-1}{2}+1}(ikz) \right],\\
   		&y_{\alpha}^{\dagger}Y^{\alpha}=\frac{\sqrt{z}}{2i}\hat{f}\left[ K_{\frac{q-1}{2}-1}(ikz)+K_{\frac{q-1}{2}+1}(ikz) \right].
   	\end{aligned}
   \end{equation}
   where the $q\times(N-q)$ matrix $\hat{f}$ should be determined by normalization conditions. Then we turn to the scalar field EOM with the rescaling $Y^{\alpha}(\nearrow)=\sqrt{z}W^{\alpha}(\nearrow)$, which reads
   \begin{equation}
   	\left[ -\hat{B}\delta^{\alpha}_{\beta}+\frac{1}{4}\delta^{\alpha}_{\beta}+\frac{q(q+1)}{8}\delta^{\alpha}_{\beta}-\frac{q+3}{2}t_i\frac{1}{2}(\sigma_i)^{\alpha}_{\ \beta} \right]W^{\beta}(\nearrow)+iy^{\alpha}(1-z\partial_z)\hat{A}^z(\nearrow)=0.
   \end{equation}
   The mixing in the gauge-flavor space is again diagonalized by the representation decomposition of $\mathfrak{su}(2)$
   \begin{equation}
   	\begin{aligned}
   		&\frac{1}{4}\delta^{\alpha}_{\beta}+\frac{q(q+1)}{8}\delta^{\alpha}_{\beta}-\frac{q+3}{2}t_i\frac{1}{2}(\sigma_i)^{\alpha}_{\ \beta}
   		\rightarrow \mathbbm{1}_q\ \oplus \ \left( \frac{q-1}{2}+1 \right)^2 \mathbbm{1}_{q-2},
   	\end{aligned}
   \end{equation}
   which leads us to use the following notation to denote the transformed fields in each irreducible representation
   \begin{equation}
   	W^{\alpha}(\nearrow)\rightarrow W_{[q]}(\nearrow)\oplus W_{[q-2]}(\nearrow); \quad y^{\alpha}\hat{f}\rightarrow y_{[q]}\hat{f}\oplus y_{[q-2]}\hat{f},
   \end{equation}
   where $W_{[q]}(\nearrow),\ y_{[q]}\hat{f} $ are $q\times(N-q)$ matrices and $W_{[q-2]}(\nearrow), y_{[q-2]}\hat{f} $ are $(q-2)\times(N-q)$ matrices. After the transformation, one finds the EOM in each irreducible representation 
   \begin{equation}
   	(-\hat{B}+1)W_{[q]}(\nearrow)+\frac{i}{4}y_{[q]}\hat{f}\left[ (2-\frac{q-1}{2})K_{\frac{q-1}{2}-1}(ikz)-(2+\frac{q-1}{2})K_{\frac{q-1}{2}+1}(ikz) \right]=0,
   \end{equation}
   \begin{equation}
   \begin{aligned}
   	&\left[-\hat{B}+(\frac{q-1}{2}+1)^2\right]W_{[q-2]}(\nearrow)\\
    &\quad\quad\quad+\frac{i}{4}y_{[q-2]}\hat{f}\left[ (2-\frac{q-1}{2})K_{\frac{q-1}{2}-1}(ikz)-(2+\frac{q-1}{2})K_{\frac{q-1}{2}+1}(ikz) \right]=0,
    \end{aligned}
   \end{equation}
   and the corresponding solutions
   \begin{equation}
   	W_{[q]}(\nearrow)=\frac{1}{2i}\frac{1}{q-1}y_{[q]}\hat{f}\left[ K_{\frac{q-1}{2}-1}(ikz)+K_{\frac{q-1}{2}+1}(ikz) \right],
   \end{equation}
   \begin{equation}
   	W_{[q-2]}(\nearrow)=\hat{h}_{[q-2]}K_{\frac{q-1}{2}+1}(ikz).
   \end{equation}
   Similarly to what was found to be the case for $W(\swarrow)_{[q+1]}$, one concludes that $W_{[q-2]}(\nearrow)$ has to be expressed in terms of only $K_{\frac{q-1}{2}+1}(ikz)$. But again, this is a natural consequence of $y_{[q-2]}=0$. One can see it similarly to before from $t_i\frac{1}{2}(\sigma_i)^{\alpha}_{\ \beta}y^{\beta}=\frac{q-2}{4}y^{\alpha}$, that is, $y^{\alpha}$ is in the representation $\pi_{q}$, which implies $y_{[q-2]}=0$. Finally, we can reorganize the solutions to read off the spectrum
    \begin{equation}
    	\begin{aligned}
    		&\hat{A}^a(\nearrow)=\frac{i}{2ik}\epsilon^{ab}k_b \hat{f} K_{\frac{q-1}{2}}(ikz),\\
    		&\hat{X}^+(\nearrow) = \hat{f} K_{\frac{q-1}{2}-1}(ikz),\\
    		&\hat{X}^-(\nearrow) =-\hat{f}K_{\frac{q-1}{2}+1}(ikz),\\
    		&Y_{[q-2]}(\nearrow)=\sqrt{z}\hat{h}_{[q-2]}K_{\frac{q-1}{2}+1}(ikz),\\
    	\end{aligned}
    \end{equation}
    where
    \begin{equation}
    	\hat{X}^{\pm}(\nearrow)=2\hat{A}^z(\nearrow)\pm \frac{i}{\sqrt{z}}y_{\alpha}^{\dagger}Y^{\alpha}(\nearrow).
    \end{equation}
    Therefore, one arrives at the spectrum in Table \ref{tab spec CB Yne}. 
    \begin{table}[htbp]
\centering
\begin{tabular}{c|c}
\multicolumn{2}{c}{$Y^{\alpha}(\nearrow)-\hat{A}^{\mu}(\nearrow)$,\ $Y_{\alpha}^{\dagger}(\swarrow)-\hat{A}^{\mu}(\swarrow)$}\\
\hline
$\nu$ &multiplicity\\
\hline
$(q-1)/2$ & $2q(N-q)$\\
\hline
$(q-1)/2-1$ & $q(N-q)$\\
\hline
$(q-1)/2+1$ & $2(q-1)(N-q)$ \\
\hline
\end{tabular}\\[10pt] 
\caption{The spectrum of the off-diagonal blocks $Y^{\alpha}(\nearrow), \hat{A}^{\mu}(\nearrow)$ and $Y_{\alpha}^{\dagger}(\swarrow)$, $\hat{A}^{\mu}(\swarrow)$. Notice that the total multiplicity adds up to $(10\, q- 4)(N-q)$ which is exactly the number of 
independent real components of the fields involved.  \label{tab spec CB Yne}}\quad\quad
\end{table}
For $q=2$ the quantity
$Y_{[q-2]}(\nearrow)$ should be set to zero but the rest
of the analysis carries through and  the multiplicities appearing in the table are  correct.
We remark that there are no unstable modes.

\subsubsection{$Y(\nwarrow)-A(\nwarrow)-\hat{A}(\nwarrow)$ mixing}

   Now we turn to the diagonal block mixing which is the most complicated part. The corresponding Lagrangian is as same as that in eq. (\ref{Lagrangian of CB}) but with every quantum part replaced with its $(\nwarrow)$ block, so we do not repeat it here.

   The idea is still to make use of decomposition of $\mathfrak{su}$(2) representations. 
   However,  it is not possible to write all classical fields in  $m_{YY}$  in a bilinear form. This means that we can no longer directly express the classical fields in terms of the $\mathfrak{su}$(2) representation matrices. However, it proves convenient to write the mixing in terms of an operator $L^{\alpha}_{\ \beta}$ defined by
   \begin{equation}
		\begin{aligned}
			L^{\alpha}_{\ \beta} X^{\beta}&=y^{\alpha}y_{\beta}^{\dagger}X^{\beta}-X^{\beta}y_{\beta}^{\dagger}y^{\alpha}=\left(\frac{1}{2}\delta^{\alpha}_{\beta}+(\sigma_i)^{\alpha}_{\ \beta}L_i\right)X^{\beta},
		\end{aligned}
	\end{equation}
   where $L_i$ is defined in eq. (\ref{def of Li}). By the $\mathfrak{su}(2)$ algebra, one finds the commutator
   \begin{equation}
		[L^{\alpha}_{\ \beta}, L^{\gamma}_{\ \theta}]=-\left( L^{\alpha}_{\ \theta}\delta^{\gamma}_{\ \beta}-L^{\gamma}_{\ \beta}\delta^{\alpha}_{\ \theta}  \right),
	\end{equation}
   and the diagonalization is still given by the  decomposition $\pi_2(s_i)\otimes(\pi_{q-1}\otimes\pi_q)(s_i) \cong \bigoplus_{r=1}^{q-1}\pi_{2r+1}\oplus\pi_{2r-1}$, which yields
   \begin{equation}
   \label{4.42}
		\begin{aligned}
			L^{\alpha}_{\ \beta}&\rightarrow\bigoplus_{r=1}^{q-1}\ r\mathbbm{1}_{2r+1}\oplus\,-r\mathbbm{1}_{2r-1}.\\
		\end{aligned}
	\end{equation}
   As the operators $L^{\gamma}_{\ \theta}L^{\theta}_{\ \gamma}$  and $L^{\alpha}_{\ \beta}$ commute they can be simultaneously diagonalized
   \begin{equation}
   \label{4.40}
   \begin{aligned}
			\delta^{\alpha}_{\ \beta}L^{\gamma}_{\ \theta}L^{\theta}_{\ \gamma}=\delta^{\alpha}_{\ \beta}(\frac{1}{2}+2L_iL_i)&\rightarrow\bigoplus_{r=1}^{q-1}2r^2\,\mathbbm{1}_{2r+1}\oplus2r^2\,\mathbbm{1}_{2r-1}.
              \end{aligned}
    \end{equation}
    Then with the operator defined above, we can write the $Y^{\dagger}-Y$ mixing in a concise form
   \begin{equation}\label{mYYdagger in terms of L}
   \begin{aligned}
       &m_{Y^{\alpha}Y_{\alpha}^{\dagger}}=\frac{1}{z^2}Y_{\alpha}^{\dagger}\delta^{\alpha}_{\ \beta}\left( \frac{1}{4}-\frac{1}{2}L^{\gamma}_{\ \theta}L^{\theta}_{\ \gamma} \right)Y^{\beta},\\
       &m_{Y^{\alpha}Y_{\beta}^{\dagger}}=\frac{1}{z^2}Y_{\alpha}^{\dagger}\left( L^{\alpha}_{\ \beta}+\frac{1}{2}L^{\alpha}_{\ \gamma}L^{\gamma}_{\ \beta}-\delta^{\alpha}_{\ \beta} \right)Y^{\beta}.
       \end{aligned}
   \end{equation}
   For the mixing $m_{YY}$, it is more useful to write its derivative with respect to $Y^{\dagger}$ which will be used in the EOM
   \begin{equation}\label{mYY in terms of L}
		\begin{aligned}
			\frac{\partial m_{YY}}{\partial Y_{\alpha}^{\dagger}}
			&=\frac{1}{2z^2}L^{\alpha}_{\ \gamma}\left( y^{\gamma}Y_{\beta}^{\dagger}y^{\beta}-y^{\beta}Y_{\beta}^{\dagger}y^{\gamma} \right)+\frac{1}{z^2}L^{\beta}_{\ \gamma}\left( y^{\alpha}Y_{\beta}^{\dagger}y^{\gamma}-y^{\gamma}Y_{\beta}^{\dagger}y^{\alpha} \right),
		\end{aligned}
	\end{equation}
   where we used the properties  given in eq.\ (\ref{2.16}) to simplify the expression. 
   It seems that we are still left with some tricky combinations of the classical fields and $Y^{\dagger}$, but we will see they are actually necessary for solving the scalar EOM. For the sake of the diagonal block mixing, one just needs to replace the quantum part with its $(\nwarrow)$ block and replace the classical solutions with their corresponding top left corners. 

   Now, as we did for the off-diagonal block mixing, we start with the gauge field EOM
   \begin{equation}\label{EOM of A}
   \begin{aligned}
		&2\left(\epsilon_{\mu\rho\nu}\partial^{\rho}-\frac{q}{z}\eta_{\mu\nu}\right)A^{\nu}(\nwarrow)\\
  &\quad\quad\quad\quad+\left(\frac{i}{z^{1/2}}\partial_{\mu}+\frac{i}{2z^{3/2}}\delta^{z}_{\mu}\right)\left( Y^{\alpha}(\nwarrow)y_{\alpha}^{\dagger}-y^{\alpha}Y_{\alpha}^{
			\dagger}(\nwarrow) \right)+\frac{2}{z}y^{\alpha}\hat{A}_{\mu}(\nwarrow)y_{\alpha}^{\dagger}=0,
   \end{aligned}
	\end{equation}
	\begin{equation}\label{EOM of Ahat}
        \begin{aligned}
		&-2\left( \epsilon_{\mu\rho\nu}\partial^{\rho}+\frac{q-1}{z}\eta_{\mu\nu} \right)\hat{A}^{\nu}(\nwarrow)\\
  &\quad\quad\quad\quad+\left( \frac{i}{z^{1/2}}\partial_{\mu}+\frac{i}{2z^{3/2}}\delta^{z}_{\mu} \right)\left( Y_{\alpha}^{\dagger}(\nwarrow)y^{\alpha}-y_{\alpha}^{\dagger}Y^{\alpha}(\nwarrow) \right)+\frac{2}{z}y_{\alpha}^{\dagger}A_{\mu}(\nwarrow)y^{\alpha}=0.
        \end{aligned}
	\end{equation}
   Then we multiply a $y^{\beta}$ to eq. (\ref{EOM of A}) from the right and to eq. (\ref{EOM of Ahat}) from the left, and add them up to gain
   \begin{equation}
   \label{EOM of Ay-yAhat}
   \begin{aligned}
		&2\bigg( \epsilon_{\mu\rho\nu}\partial^{\rho}\delta^{\alpha}_{\beta}+\frac{1}{z}\eta_{\mu\nu} L^{\alpha}_{\ \beta}\bigg) \bigg( A^{\nu}(\nwarrow)y^{\beta}-y^{\beta}\hat{A}^{\nu}(\nwarrow) \bigg)\\
  &\quad\quad+\bigg( \frac{i}{z^{1/2}}\partial_{\mu}+\frac{i}{2z^{3/2}}\delta^{z}_{\mu} \bigg)\bigg( -L^{\alpha}_{\ \beta}Y^{\beta}(\nwarrow)+y^{\alpha}Y_{\beta}^{\dagger}(\nwarrow)y^{\beta}-y^{\beta}Y_{\beta}^{\dagger}(\nwarrow)y^{\alpha} \bigg)=0.
  \end{aligned}
	\end{equation}
   This equation is almost the same as that in eq.\ (\ref{massive CS EOM}) but with mixing in the 
    gauge and R-symmetry indices specified by the operator $L^{\alpha}_{\ \beta}$, which should again be diagonalized by the representation decomposition
   \begin{equation}
       \pi_2\otimes\pi_{q-1}\otimes\pi_q\cong\bigoplus_{r=1}^{q-1}\pi_{2r+1}\oplus\pi_{2r-1}\cong\bigoplus_{\boldsymbol{\ell}=1}^{q-1}\boldsymbol{\ell}\oplus\boldsymbol{\ell-1},
   \end{equation}
   where the bold symbol $\boldsymbol{\ell}$ denotes the spin $\ell$ representation. The states of the deduced irreducible representations are found to be $y^{\alpha} \hat{Y}_{\ell}^{m}, Y_{\ell-1}^{m} y^{\alpha}$ with $\ell=1,\dots,q-1$, which fulfill
   \begin{equation}
       \begin{aligned}
           &(J_3)^{\alpha}_{\ \beta}\,y^{\beta}\hat{Y}_{\ell}^{m}=my^{\alpha}\hat{Y}_{\ell}^{m}, \quad (J_i)^{\alpha}_{\ \beta}(J_i)^{\beta}_{\ \gamma}\,y^{\gamma}\hat{Y}_{\ell}^{m}=\ell(\ell+1)y^{\alpha}\hat{Y}_{\ell}^{m},\\
           &(J_3)^{\alpha}_{\ \beta} \,Y_{\ell}^{m}y^{\beta}=mY_{\ell}^{m}y^{\alpha}, \quad (J_i)^{\alpha}_{\ \beta}(J_i)^{\beta}_{\ \gamma}\,Y_{\ell}^{m}y^{\gamma}=\ell(\ell+1)Y_{\ell}^{m}y^{\alpha}.\\
       \end{aligned}
   \end{equation}
   Note that $Y_{\ell}^{m}$ is the fuzzy spherical harmonics with respect to $t_i$. We refer to Appendix \ref{Mfuzzy} for a more detailed discussion of these states. Then, we for convenience define 
   \begin{equation}\label{shorthands for CB DB}
       B^{\mu,\alpha}(\nwarrow)=A^{\mu}(\nwarrow)y^{\alpha}-y^{\alpha}\hat{A}^{\mu}(\nwarrow), \quad \sqrt{z}V^{\alpha}(\nwarrow)=y^{\alpha}Y_{\beta}^{\dagger}(\nwarrow)y^{\beta}-y^{\beta}Y_{\beta}^{\dagger}(\nwarrow)y^{\alpha}.
   \end{equation}
   which can be further expressed in terms of the states we constructed above:
   \begin{equation}
       B^{\mu,\alpha}(\nwarrow)=B^{\mu}_{\ell, m} y^{\alpha}\hat{Y}_{\ell}^{m}+B_{-\ell, m}^{\mu}Y_{\ell-1}^{m}y^{\alpha}, \quad V^{\alpha}(\nwarrow)=V_{\ell, m}\,y^{\alpha}\hat{Y}_{\ell}^{m}+V_{-\ell,m}Y_{\ell-1}^{m}y^{\alpha}.
   \end{equation}
   Similarly, we write
   \begin{equation}
       Y^{\alpha}(\nwarrow)=\mathcal{Y}_{\ell m}\,y^{\alpha}\hat{Y}_{\ell}^{m}+\mathcal{Y}_{-\ell,m}Y_{\ell-1}^{m}y^{\alpha},
   \end{equation}
   where the subscript $-\ell$ of the component attached to $Y_{\ell-1}^{m}y^{\alpha}$ is made for the sake of a concise expression in what follows.

   Now, plugging the above expression into the EOM (\ref{EOM of Ay-yAhat}), one finds
   \begin{equation}
       \begin{aligned}
           &2\bigg( \epsilon_{\mu\rho\nu}\partial^{\rho}\pm\frac{\ell}{z}\eta_{\mu\nu}\bigg)B_{\pm\ell, m}^{\nu}+\bigg( \frac{i}{z^{1/2}}\partial_{\mu}+\frac{i}{2z^{3/2}}\delta^{z}_{\mu} \bigg)\left( \mp \ell Y_{\pm\ell, m}+\sqrt{z}V_{\pm\ell, m}\right)=0.\\
       \end{aligned}
   \end{equation}
   By comparing the above equations with eq. (\ref{massive CS EOM}), one finds the solutions
   \begin{equation}\label{sol to gauge field EOM}
		\begin{aligned}
			B_{\pm\ell,m}^{a}(\nwarrow)&=-\frac{i}{2ik}\epsilon^{ab}k_{b}f_{\pm \ell,m} K_{\ell}(ikz),\\
			B_{\pm\ell,m}^z(\nwarrow)&=\mp\frac{1}{4}f_{\pm\ell,m}\left(K_{\ell-1}(ikz)-K_{\ell+1}(ikz)\right),\\
			\mp \ell Y_{\pm\ell,m}(\nwarrow)+\sqrt{z}V_{\pm\ell,m}(\nwarrow)&=-\frac{\sqrt{z}}{2i}f_{\pm\ell,m}\left(K_{\ell-1}(ikz)+K_{\ell+1}(ikz)\right).
		\end{aligned}
	\end{equation}
   where $f_{\pm\ell,m}$ should be determined by normalization. However, for a fixed Lorentz index $\mu$, we solved only $2q(q-1)$ components of the recombined field $B^{\mu,\alpha}(\nwarrow)$, while there are $q^2+(q-1)^2$ DOF's in $A^{\mu},\hat{A}^{\mu}$, which implies that we are still left with one unidentified DOF of the gauge fields. It turns out that this one DOF is encoded in $\text{tr}(A^{\mu})-\text{tr}(\hat{A}^{\mu})$ which evaluates according to the following equation
   \begin{equation}
		\epsilon_{\mu\rho\nu}\partial^{\rho}\left( \text{tr}(A^{\nu}(\nwarrow))-\text{tr}(\hat{A}^{\nu}(\nwarrow)) \right)=0,
	\end{equation}
    where we trace eq. (\ref{EOM of A}), (\ref{EOM of Ahat}) on both sides and subtract the resulting equations. Thus, the remaining single DOF is a free massless Chern-Simons gauge field. Hereby we have solved all DOF's of the gauge fields.

    Next, we consider the scalar EOM
    \begin{equation}
    	\begin{aligned}
    		&\partial_{\mu}\partial^{\mu}Y^{\alpha}(\nwarrow)-\frac{1}{z^2}\left( -\frac{3}{4}\delta^{\alpha}_{\ \beta}-\frac{1}{2}\delta^{\alpha}_{\ \beta}L^{\gamma}_{\ \theta}L^{\theta}_{\ \gamma}+L^{\alpha}_{\ \beta}+\frac{1}{2}L^{\alpha}_{\ \gamma}L^{\gamma}_{\ \beta} \right)Y^{\beta}(\nwarrow)\\
    		&-\frac{1}{2z^2}L^{\alpha}_{\ \gamma}\left( y^{\gamma}Y_{\beta}^{\dagger}(\nwarrow)y^{\beta}-y^{\beta}Y_{\beta}^{\dagger}(\nwarrow)y^{\gamma} \right)-\frac{1}{z^2}L^{\beta}_{\ \gamma}\left( y^{\alpha}Y_{\beta}^{\dagger}(\nwarrow)y^{\gamma}-y^{\gamma}Y_{\beta}^{\dagger}(\nwarrow)y^{\alpha} \right)\\
    		&-i\left( \frac{1}{z^{3/2}}\delta^{z}_{\mu}-\frac{1}{z^{1/2}}\partial_{\mu} \right)\left( A^{\mu}(\nwarrow)y^{\alpha}-y^{\alpha}\hat{A}^{\mu}(\nwarrow) \right)=0,
    	\end{aligned}
    \end{equation}
    where we have used eqs. (\ref{mYYdagger in terms of L}) and (\ref{mYY in terms of L}) to rewrite the mixing terms. With the rescaling $Y^{\alpha}(\nwarrow)=\sqrt{z}W^{\alpha}(\nwarrow)$, one further finds
    \begin{equation}\label{CB EOM scalar W}
    	\begin{aligned}
    		&\left[ -\hat{B}\delta^{\alpha}_{\ \beta}+\left( \delta^{\alpha}_{\ \beta}+\frac{1}{2}\delta^{\alpha}_{\ \beta}L^{\gamma}_{\ \theta}L^{\theta}_{\ \gamma}-L^{\alpha}_{\ \beta}-L^{\alpha}_{\ \gamma}L^{\gamma}_{\ \beta} \right) \right]W^{\beta}(\nwarrow)\\
            &-\frac{1}{2}L^{\alpha}_{\ \gamma}\bigg( -L^{\gamma}_{\ \beta}W^{\beta}(\nwarrow)+V^{\gamma}(\nwarrow) \bigg)\\
            &-L^{\beta}_{\ \gamma}\left( y^{\alpha}W_{\beta}^{\dagger}(\nwarrow)y^{\gamma}-y^{\gamma}W_{\beta}^{\dagger}(\nwarrow)y^{\alpha} \right)
    		-i\left( \delta^{z}_{\mu}-z\partial_{\mu} \right)B^{\mu,\alpha}(\nwarrow)=0.
    	\end{aligned}
    \end{equation}
    It is not possible to write the first term in the third line as the action of $L^{\alpha}_{\ \beta}$ on a variable defined above. However, transforming to the basis that diagonalizes $L^{\alpha}_{\ \beta}$ we find a useful formula
    \begin{equation}\label{eigen formula}
	    	\left(L^{\beta}_{\ \gamma}\left( y^{\alpha}W_{\beta}^{\dagger}y^{\gamma}-y^{\gamma}W_{\beta}^{\dagger}y^{\alpha} \right)\right)_{\pm\ell,m}=(\mp \ell+1)V_{\pm\ell,m},
    \end{equation}
    which enables us to keep only $W(\nwarrow)$ in eq. (\ref{CB EOM scalar W}) with the other fields replaced with the explicit expression by diagonalizing $L^{\alpha}_{\ \beta}$ and using the solutions to the gauge field EOM. Then using eq. (\ref{4.40}) the scalar EOM  simplifies to 
    \begin{equation}
    	\left[ -\hat{B}+(\ell\mp1)^2 \right]W_{\pm\ell,m}-if_{\pm\ell,m}K_{\ell\pm1}(ikz)=0,
    \end{equation}
    which seems to require $W_{\pm\ell,m}$ to be given in terms of $K_{\ell\pm1}(ikz)$. But as the term in the square bracket cancels $K_{\ell\mp1}(ikz)$ in the $\boldsymbol{\pm\ell}$ representation, one is free to add such Bessel functions to the corresponding solutions. Therefore, the generic solution reads
    \begin{equation}
        W_{\pm\ell,m}=\mp\frac{i}{4\ell}f_{\pm\ell,m}K_{\ell\pm1}(ikz)+g_{\pm\ell,m}K_{\ell\mp1}(ikz),
    \end{equation}
    where $g_{\pm\ell,m}$ is independent of $f_{\pm\ell,m}$ and  should also be determined by normalization.

    Now, as all DOF's in the diagonal block mixing are accounted for, we can reorganize the solutions to read off the spectrum
    \begin{equation}\label{sols of CB DB diagonalized}
    	\begin{aligned}
    	&2B^z_{\pm\ell,m}\mp i\left( 3\ell W_{\pm\ell,m}\pm V_{\pm\ell,m} \right)=\mp i4\ell g_{\pm\ell,m}K_{\ell\mp1}(ikz),\\
            &2B_{\ell,m}^z\mp i\left( -\ell W_{\ell,m}+V_{\ell,m} \right)=\pm f_{\ell,m}K_{\ell\pm1}(ikz),\\
            &2B_{-\ell,m}^z\pm i\left( \ell W_{-\ell,m}+V_{-\ell,m} \right)=\mp f_{-\ell,m}K_{\ell\pm1}(ikz),
    	\end{aligned}
    \end{equation}
    where we also organize them in such way that it is easy to find the combinations linearly independent of each other. Together with $B^{\mu,\alpha}(\nwarrow)$ and $\text{tr}(A^{\mu}(\nwarrow))-\text{tr}(\hat{A}^{\mu}(\nwarrow))$, we now found a number of linearly independent solutions which match the number of DOF of the fields.

\begin{table}[htbp]
\centering
\begin{tabular}{c|c}
\multicolumn{2}{c}{$Y^{\alpha}(\nwarrow)-Y_{\alpha}^{\dagger}(\nwarrow)-A^{\mu}(\nwarrow)-\hat{A}^{\mu}(\nwarrow)$}\\
\hline
$\nu$ &multiplicity\\
\hline
$\ell-1$ & $6\ell+1$\\
\hline
$\ell+1$ & $6\ell-1$\\
\hline
$\ell$ & $8\ell$ \\
\hline
1/2 (massless) & $3$\\
\hline
\end{tabular}\\[10pt] 
\caption{The spectrum of the diagonal blocks $Y^{\alpha}(\nwarrow),Y_{\alpha}(\nwarrow),A^{\mu}(\nwarrow), \hat{A}^{\mu}(\nwarrow)$, where $\ell=1,2,\dots,q-1$.  Notice that the total multiplicity adds up to $10\, q(q-1)+3$ which exactly matches number of independent, real
field components of the fields involved. \label{tab spec CB Ynw}}\quad\quad
\end{table}

For the $Y^{\alpha}(\searrow)$ and $Y_{\alpha}^{\dagger}(\searrow)$, we have remaining $4(N-q)(N-q+1)$ massless DOF. Similarly, for $A^{\mu}(\searrow)$ and $\hat{A}^{\mu}(\searrow)$, we have $3(N-q)(N-q)$ and $3(N-q+1)(N-q+1)$ massless DOF respectively.

\section{Summary and conclusion\label{conclusion}}

We have considered ABJM theory with a non-trivial vacuum which corresponds to the presence of a 1/2-BPS domain wall, and we have diagonalized
the quadratic part of the corresponding quantum action, i.e. we have determined the spectrum of its quantum
fields.  We summarize the spectrum in the two tables below. Table~\ref{finaleasy} shows the spectrum of the easy fields (and is 
an extension of table~\ref{tab spec EF}), whereas table~\ref{finalcomplicated}
summarizes the spectrum of the complicated fields. It should be noted, though, that to obtain the structure
below we have joined $2q(q-1)$ excitations from the diagonal part of the complicated fields to the list of easy fields. Furthermore, we have chosen
a convention where in order to
count all degrees of freedom the multiplicities in 
the table should be multiplied by two for complex fields, i.e.\ for $Y$'s 
and $\psi$'s.\footnote{ 
The fields in the fourth columns should be understood as recombined objects, 
such as those appearing in eq. (\ref{sols of CB DB diagonalized}), but for simplicity we still use the notation of original fields. }

\begin{table}[htbp]
\centering
    \begin{equation}
    	\begin{tabular}{|c|c|c|c|}
    		\hline
    		Multiplicity& $\nu$ of $Y^{\tilde{\alpha}}$ & $m$ of $\psi_{\alpha}$ & $\nu$ of $Y^{\alpha}$ \\ \hline
    		$2\ell+1$ & $\ell$ & $-\ell+\frac{1}{2}$ & $\ell-1$ \\[3pt]
    		$2\ell-1$ & $\ell$ & $\ell+\frac{1}{2}$ & $\ell+1$ \\[3pt] \hline
            Multiplicity & $\nu$ of $Y^{\tilde{\alpha}}$ & $m$ of $\psi_{\alpha}$ & $\nu$ of $A^z$ \\
             \hline
    		$(q+1)(N-q+1)$ & $\frac{q}{2}$ &  $-\frac{q}{2}+\frac{1}{2}$ & $\times$\\[3pt] 
    		$(q-1)(N-q+1)$ & $\frac{q}{2}$ & $\frac{q}{2}+\frac{1}{2}$ & $\frac{q}{2}+1$ \\[3pt] \hline
            Multiplicity & $\nu$ of $Y^{\tilde{\alpha}}$ & $m$ of $\psi_{\alpha}$ & $\nu$ of $\hat{A}^z$ \\
             \hline
    		$q(N-q)$ & $\frac{q-1}{2}$ & $-\frac{q-1}{2}+\frac{1}{2}$ & $\frac{q-1}{2}-1$ \\[3pt] 
    		$(q-2)(N-q)$ & $\frac{q-1}{2}$ & $\frac{q-1}{2}+\frac{1}{2}$ & $\times$ \\ \hline
            Multiplicity& $\nu$ of $Y^{\tilde{\alpha}}$ & $m$ of $\psi_{\alpha}$ & $\nu$ of $Y^{\alpha}$ 
            \\ \hline
    		$2(N-q)(N-q+1)$ & $\frac{1}{2}$ & 0 & $\frac{1}{2}$ \\ \hline
    	\end{tabular}
    \end{equation}
    \caption{ \label{finaleasy} The spectrum of easy fields with $\ell=1,\ldots q-1$. The notation $\times$ means that the field does not have any mass with the given multiplicity.}
    \end{table}
\begin{table}[htbp]
\centering
\begin{equation}
		\begin{tabular}{|c|c|c|c|}
			\hline 
			Multiplicity& \text{$\nu$ of $B^{0,\alpha},B^{1,\alpha}$}  & \text{$m$ of $\psi_{\tilde{\alpha}}$} & \text{$\nu$ of $B^{z,\alpha}$}  \\  \hline
			$2\ell$ & $\ell$ & $\ell-\frac{1}{2}$ & $\ell-1$ \\[3pt]
			$2\ell$ & $\ell$ & $-\ell-\frac{1}{2}$ & $\ell+1$  \\
            $1$ & $\frac{1}{2}$ & $\times$ & $\frac{1}{2}$
            \\[3pt] \hline
			Multiplicity & $\nu$ of $A^0,A^1$ & $m$ of $\psi_3,\psi_4$ & $\nu$ of $Y^1, Y^2$  \\ \hline 
			$(q-1)(N-q+1)$ & $\frac{q}{2}$ & $\frac{q}{2}-\frac{1}{2}$ & $\frac{q}{2}-1$ \\
            $(N-q+1)$ & $\times$ & $\frac{q}{2}-\frac{1}{2}$ & $\frac{q}{2}-1$
            \\[3pt] \hline 
			Multiplicity & $\nu$ of $\hat{A}^0,\hat{A}^1$ & $m$ of $\psi_3,\psi_4$ & $\nu$ of $Y^1, Y^2$  \\ \hline
			$(q-1)(N-q)$ & $\frac{q-1}{2}$ & $-\frac{q}{2}$ & $\frac{q-1}{2}+1$ \\
            $(N-q)$ & $\frac{q-1}{2}$ & $\times$ & $\times$
            \\[3pt] \hline 
                Multiplicity & $\nu$ of $B^{0,\alpha},B^{1,\alpha}$ & $m$ of $\psi_{\tilde{\alpha}}$ & $\nu$ of $B^{z,\alpha}$  \\ \hline 
			$2(N-q)(N-q+1)$ & $\frac{1}{2}$ & $0$ & $\frac{1}{2}$ \\
            $1$ & $\frac{1}{2}$ & $\times$ & $\frac{1}{2}$
            \\[3pt] \hline 
		\end{tabular}
	\end{equation}
	\caption{\label{finalcomplicated}The spectrum of complicated fields where $\ell=1,\ldots q-1$.We recall that $B^{\mu,\alpha}=A^{\mu}y^{\alpha}-y^{\alpha}\hat{A}^{\mu}$, which was introduced in eq. (\ref{shorthands for CB DB}) for the $(\nwarrow)$ block. Thus, the masses (or $\nu$) of $B^{\mu}$ should be understood as those of the pair $(A^{\mu},\hat{A}^{\mu})$.}
	\end{table}

The spectrum shows the characteristics of a supersymmetric field theory, namely that  the fields can be combined into multiplets with a
difference in conformal dimension of $\pm 1/2$ between fermions and bosons.
 A similar pattern was found in the quantization around the ${\cal N}=4$ SYM vacuum dual to the D3-D5 probe brane system with flux which is likewise 1/2-BPS~\cite{Buhl-Mortensen:2016pxs,Buhl-Mortensen:2016jqo}. This should be contrasted with the case of  non-supersymmetric domain walls
in ${\cal N}=4$ SYM dual to D3-D7 probe brane systems, where such a structure was not observed~\cite{GimenezGrau:2018jyp,Gimenez-Grau:2019fld}. 
 
 Knowing the spectrum of the quantum fields  the perturbative program can easily be set up. For the excitations whose
spectrum was found by direct diagonalization (i.e. fermions and easy bosons) we can immediately read off the propagators from
 eq.~(\ref{eq:PropagatorBessel}), and for the excitations whose masses were found by solution of the appropriate equations of motion we can obtain the
 propagators by pasting together solutions with the appropriate boundary conditions and normalization which again will lead to 
 an expression of the same type as (the last line in)  eq.~(\ref{eq:PropagatorBessel}).
 
 This opens up the possibility of perturbatively computing the expectation values of local as well as non-local observables in the defect 
 background, such as correlation functions of local operators and Wilson loops. The 1/2-BPS domain wall has been shown to correspond
 to an integrable boundary state and accordingly a closed expression for tree-level one-point functions has been derived~\cite{Gombor:2021hmj,Kristjansen:2021abc,Gombor:2022aqj}. A perturbative computation of one-point functions at higher
loop orders could provide input for the derivation of a possible asymptotic all loop expression for one-point functions based on 
integrability bootstrap in the spirit of~\cite{Gombor:2020kgu,Komatsu:2020sup,Gombor:2020auk,Gombor:2024api}. Furthermore, the perturbative computation of lower point correlation functions could form the starting point for the application  of the boundary conformal
bootstrap program to the 1/2 BPS domain wall  in ABJM theory.

Finally, let us mention that the techniques developed for the computation of one-point functions in the domain wall version of ABJM theory work equally well for the computation of one-point functions of mass deformed ABJM theory. This holds both for the integrability based methods applied at tree level in~\cite{Kristjansen:2021abc,Gombor:2022aqj} and the perturbative approach followed in the present paper for higher loop computations.

\acknowledgments

C.K.\ was supported by DFF-FNU through grant number 1026-00103B. X.Q. was supported by CSC through grant number
20220794001 and C.S was supported by GEP through grant number JY202211.
We thank J.\ Ambj\o rn, E. Leeb-Lundberg and K.\ Zarembo for useful discussions.

\appendix

\section{Modified fuzzy spherical harmonics\label{Mfuzzy}}

In this section we explicitly construct the modified fuzzy spherical harmonics which furnish the $\pi_{q-1}\otimes\pi_q$ representation of $\mathfrak{su}(2)$, and then generalize it to a spin $1/2$ version which forms the $\pi_2\otimes\pi_{q-1}\otimes\pi_q$ representation. We start by recalling a few basic facts about
$\mathfrak{su}(2)$ representation theory.

\subsection{Elements of $\mathfrak{su}(2)$ representation theory }

    Letting $\pi_p$ denote the $p$-dimensional irreducible representation of su(2), we have the following decomposition formula 
    \begin{equation}
    \pi_p\otimes\pi_q\cong\pi_{p+q-1}\oplus\pi_{p+q-3}\oplus\cdots\oplus\pi_{|p-q|+1}, 
    \end{equation}
    which implies that the Casmir of $\pi_p\otimes\pi_q$, denoted by $\mathcal{C}_2^{p\otimes q}$, can be diagonalized as
        \begin{equation}
			\mathcal{C}_2^{p\otimes q}=\bigoplus_{r=1}^{q-1} \mathcal{C}_2^{2r}
			=\bigoplus_{r=1}^{q-1}\frac{(2r-1)(2r+1)}{4}\mathbbm{1}_{2r},
		\end{equation}
    where it is known that 
    \begin{equation}
        \mathcal{C}_2^{p}=\frac{(p+1)(p-1)}{4}.
    \end{equation}
    Another Casimir-like operator is $\pi_p(s_i)\otimes\pi_q(s_i)$ where $s_i$ denotes $\mathfrak{su}(2)$ elements with $i=1,2,3$ and the repeated $i$'s are to be summed over. It is diagonalized as 
    \begin{equation}
    \begin{aligned}
		\pi_p(s_i)\otimes\pi_q(s_i)&=\frac{1}{2} \left( \mathcal{C}_2^{p\otimes q} - \mathcal{C}_2^p\otimes\mathbbm{1}_q - \mathbbm{1}_p\otimes \mathcal{C}_2^q\right)\\
        &=\bigoplus_{r=1}^{q-1}\left(\frac{(2r-1)(2r+1)}{4}-\frac{(p+1)(p-1)}{4}-\frac{(q+1)(q-1)}{4}\right)\mathbbm{1}_{2r}.
    \end{aligned}
    \end{equation}

\subsection{Modified fuzzy spherical harmonics for $\pi_{q-1}\otimes\pi_q$}

The starting point are the standard fuzzy spherical harmonics~\cite{Hoppe82,deWit88}.
The standard fuzzy sphere harmonics $Y^{m}_{\ell}(\hat{t})$ for the $q$-dimensional irrep fulfill
\begin{equation}
    [\hat{t}_3,Y^{m}_{\ell}(\hat{t})]=m Y^{m}_{\ell}(\hat{t}), \quad \ell=0,1,\dots,q-1,
\end{equation}
which motivates us to define $\hat{Y}^{m}_{\ell}=Y^{m}_{\ell}(\hat{t})^T$ such that
\begin{equation}
    [-\hat{t}_3^T,\hat{Y}_{\ell}^{m}]=m \hat{Y}^{m}_{\ell},
\end{equation}
while the action of the ladder operators reads
\begin{equation}\label{action of ladder operators}
\begin{aligned}
    &[-\hat{t}_+^T,\hat{Y}_{\ell}^{m}]=\sqrt{(\ell-m)(\ell+m+1)} \hat{Y}^{m+1}_{\ell},\\
    &[-\hat{t}_-^T,\hat{Y}_{\ell}^{m}]=\sqrt{(\ell+m)(\ell-m+1)} \hat{Y}^{m-1}_{\ell}.
\end{aligned}
\end{equation}
By the BPS equations, one finds the following formulas for $L_i$ which are defined by eq. (\ref{def of Li})
\begin{equation}\label{action of L on y}
   \begin{aligned}
    &L_3y^1=-\frac{1}{2}y^1, \quad L_+ y^1=-y^2, \quad L_- y^1=0,\\
    &L_3y^2=\frac{1}{2}y^1, \quad L_+ y^2=0, \quad L_- y^2=-y^1,\\
    \end{aligned}
\end{equation}
where $L_{\pm}=L_1\pm iL_2$. Then noticing that ($i=1,2,3,\pm$ and $\alpha=1,2$)
\begin{equation}
    L_i y^{\alpha}\hat{Y}_{\ell}^{m}=(L_i y^{\alpha})\hat{Y}_{\ell}^{m}+y^{\alpha}[-\hat{t}_i^T,\hat{Y}_{\ell}^{m}],
\end{equation}
one  easily finds 
\begin{equation}\label{L on modified fuzzy sphere harmonics}
\begin{aligned}
    &L_3\, y^1 \hat{Y}^{m}_{\ell}=(m-\frac{1}{2})y^1 \hat{Y}^{m}_{\ell},\quad L_+y^1\hat{Y}_{\ell}^{\ell}=-y^2\hat{Y}_{\ell}^{\ell}, \quad L_-\, y^1 \hat{Y}^{-\ell}_{\ell}=0,\\
    &L_3\, y^2 \hat{Y}^{m}_{\ell}=(m+\frac{1}{2})y^2 \hat{Y}^{m}_{\ell}, \quad L_+\, y^2 \hat{Y}^{\ell}_{\ell}=0,\quad L_-\, y^2 \hat{Y}^{-\ell}_{\ell}=-y^1 \hat{Y}_{\ell}^{-\ell},
\end{aligned}
\end{equation}
Now, by writting the Casimir in terms of $L_3$ and the ladder operators
\begin{equation}
    L_i L_i = L_3^2-L_3+L_+L_-= L_3^2+L_3+L_-L_+,
\end{equation}
one further obtains
\begin{equation}
    L_i L_i\, y^1 \hat{Y}^{-\ell}_{\ell}=(\ell+\frac{1}{2})(\ell+\frac{3}{2})y^1 \hat{Y}^{-\ell}_{\ell},\quad L_i L_i\, y^2 \hat{Y}^{\ell}_{\ell}=(\ell+\frac{1}{2})(\ell+\frac{3}{2})y^2 \hat{Y}^{\ell}_{\ell},
\end{equation}
which implies that $y^1 \hat{Y}^{-\ell}_{\ell}$ is the lowest state in the $2(\ell+1)$-dimensional irrep, while $y^2 \hat{Y}^{\ell}_{\ell}$ is the highest state in the $2(\ell+1)$-dimensional irrep. Thus, $y^2\hat{Y}^{\ell}_{\ell}$ ($y^1\hat{Y}^{-\ell}_{\ell}$) with $\ell=0,1,\dots,q-2$ include the highest (lowest) states of all the irreps deduced from $\pi_{q}\otimes\pi_{q-1}$, and we can obtain the entire set of states in each irrep by acting on $y^2\hat{Y}^{\ell}_{\ell}$ ($y^1\hat{Y}^{-\ell}_{\ell}$) with $L_-$ ($L_+$). The states constructed in the two ways are respectively given by $(n=0,1,\dots,2\ell+1)$
\begin{equation}
    \begin{aligned}
        (L_-)^{n}y^2\hat{Y}_{\ell}^{\ell}=\sqrt{\frac{(2\ell)!(n-1)!}{(2\ell-(n-1))!}}\left( -n y^1\hat{Y}_{\ell}^{\ell-(n-1)}+\sqrt{(2\ell-n+1)n}\,y^2\hat{Y}_{\ell}^{\ell-n} \right),
    \end{aligned}
\end{equation}
\begin{equation}
    \begin{aligned}
        (L_+)^{n}y^1\hat{Y}_{\ell}^{-\ell}=\sqrt{\frac{(2\ell)!(n-1)!}{(2\ell-(n-1))!}}\left( -n y^2\hat{Y}_{\ell}^{-\ell+(n-1)}+\sqrt{(2\ell-n+1)n}\,y^1\hat{Y}_{\ell}^{-\ell+n} \right).
    \end{aligned}
\end{equation}
One easily finds $(L_-)^{n}y^2\hat{Y}_{\ell}^{\ell}\propto (L_+)^{2(\ell+1)-n}y^1\hat{Y}_{\ell}^{-\ell}$ which verifies the consistency of the two ways of generating states. Hence, up to a normalization factor, the states in the spin $\ell+1/2$ irrep can be expressed as
\begin{equation}
    T_{\ell+1/2}^{m+1/2}=-\sqrt{l-m}\,y^1\hat{Y}_{\ell}^{m+1}+\sqrt{\ell+m+1}\,y^2\hat{Y}_{\ell}^{m},
\end{equation}
where $\ell=0,1,\dots,q-2$ and $m=-\ell-1,-\ell,\dots
,\ell$. We also assume $\hat{Y}_{\ell}^{-\ell-1}=\hat{Y}_{\ell}^{\ell+1}=0$. These states carry different eigenvalues of the Casimir and $L_3$ so they are automatically orthogonal to each other.

\subsection{Spin $1/2$ modified fuzzy spherical harmonics for $\pi_2\otimes\pi_{q-1}\otimes \pi_q$}

As the mixing of the complicated field is governed by the $\mathfrak{su}(2)$
representation $\pi_2\otimes\pi_{q-1}\otimes \pi_{q}$, which is specified by the operator
\begin{equation}
    (J_i)^{\alpha}_{\ \beta}=\frac{1}{2}(\sigma_i)^{\alpha}_{\ \beta}+\delta^{\alpha}_{\ \beta}L_i,
\end{equation}
we also need to construct the states of the irreps deduced from it. Then, one by eq. (\ref{L on modified fuzzy sphere harmonics}) finds
\begin{equation}
    (J_3)^{\alpha}_{\ \beta}\,y^{\beta}\hat{Y}_{\ell}^{m}=my^{\alpha}\hat{Y}_{\ell}^{m}, \quad (J_+)^{\alpha}_{\ \beta}\,y^{\beta}\hat{Y}_{\ell}^{\ell}=(J_-)^{\alpha}_{\ \beta}\,y^{\beta}\hat{Y}_{\ell}^{-\ell}=0,
\end{equation}
where 
\begin{equation}
    (J_{\pm})^{\alpha}_{\ \beta}=(J_1)^{\alpha}_{\ \beta}\pm i (J_2)^{\alpha}_{\ \beta}.
\end{equation}
Hence, again with a rewriting of the Casimir
\begin{equation}
\begin{aligned}
    (J_i)^{\alpha}_{\ \gamma}(J_i)^{\gamma}_{\ \beta}&=(J_3)^{\alpha}_{\ \gamma}(J_3)^{\gamma}_{\ \beta}-(J_3)^{\alpha}_{\ \beta}+(J_+)^{\alpha}_{\ \gamma}(J_-)^{\gamma}_{\ \beta}\\
    &=(J_3)^{\alpha}_{\ \gamma}(J_3)^{\gamma}_{\ \beta}+(J_3)^{\alpha}_{\ \beta}+(J_-)^{\alpha}_{\ \gamma}(J_+)^{\gamma}_{\ \beta},
\end{aligned}
\end{equation}
one further finds
\begin{equation}
    (J_i)^{\alpha}_{\ \gamma}(J_i)^{\gamma}_{\ \beta}\,y^{\beta}\hat{Y}_{\ell}^{\ell}=\ell(\ell+1)y^{\alpha}\hat{Y}_{\ell}^{\ell}, \quad (J_i)^{\alpha}_{\ \gamma}(J_i)^{\gamma}_{\ \beta}\,y^{\beta}\hat{Y}_{\ell}^{-\ell}=\ell(\ell+1)y^{\alpha}\hat{Y}_{\ell}^{-\ell},
\end{equation}
which implies $y^{\alpha}\hat{Y}_{\ell}^{\ell}$ ($y^{\alpha}\hat{Y}_{\ell}^{-\ell}$) is the highest (lowest) state in the irrep of spin $\ell$. Meanwhile, one can construct the state $Y_{\ell}^{m}y^{\alpha}$, where $Y_{\ell}^{m}$ is the fuzzy sphere harmonics in terms of $t_i$ which fulfills
\begin{equation}
    [t_3, Y_{\ell}^{m}]=m Y_{\ell}^{m}.
\end{equation}
It is not hard to verify that the above action of $J_i$'s also holds for $Y_{\ell}^{m}y^{\alpha}$ but with $\ell=0,1,\dots,q-2$ (as $t_i$ is the $(q-1)$-dimensional representation). One also notes that $Y_{\ell}^{m}y^{\alpha}$ is linearly independent of $y^{\alpha}\hat{Y}_{\ell}^{m}$, i.e. they are not proportional to each other. 

Now, with the representation decomposition
\begin{equation}
    \frac{1}{2}\otimes \frac{q-2}{2}\otimes\frac{q-1}{2}=\bigoplus_{\ell=1}^{q-1}\boldsymbol{\ell}\oplus\boldsymbol{(\ell-1)},
\end{equation}
where $\boldsymbol{\ell}$ denotes the spin $\ell$ representation of su(2), one finds the highest state of each irrep (note that $Y_{0}^{0}=\mathbbm{1}_{q-1}$)
\begin{equation}
    \boldsymbol{\ell}: y^{\alpha}\hat{Y}_{\ell}^{\ell}, \quad \boldsymbol{\ell-1}: Y_{\ell-1}^{\ell}y^{\alpha}, \quad \ell=1,2,\dots,q-1.
\end{equation}
Then, the entire set of states of each irrep. can again be generated by acting on the highest state with $J_-$ (or the other way around). A straightforward calculation shows that the descendant states have a simple form
\begin{equation}
    (J_-)^{\alpha}_{\ \beta_1}\cdots(J_-)^{\beta_{n-2}}_{\ \beta_{n-1}}(J_-)^{\beta_{n-1}}_{\ \beta_{n}}y^{\alpha}\hat{Y}_{\ell}^{\ell}\propto y^{\alpha}\hat{Y}_{\ell}^{\ell-n},
\end{equation}
\begin{equation}
    (J_-)^{\alpha}_{\ \beta_1}\cdots(J_-)^{\beta_{n-2}}_{\ \beta_{n-1}}(J_-)^{\beta_{n-1}}_{\ \beta_{n}}Y_{\ell}^{\ell}y^{\alpha}\propto Y_{\ell}^{\ell-n}y^{\alpha}.
\end{equation}
Therefore, the entire set of states of each irreducible representation is given by
\begin{equation}
\begin{aligned}
    &\boldsymbol{\ell}: y^{\alpha}\hat{Y}_{\ell}^{m}, \quad m=-\ell,-\ell+1,\dots,\ell,\\
    &\boldsymbol{\ell-1}: Y_{\ell-1}^{m}y^{\alpha}, \quad m=-\ell+1,-\ell+2,\dots,\ell-1.
\end{aligned}
\end{equation}
Again, these states are not yet normalized, and $y^{\alpha}\hat{Y}_{\ell}^{m},Y_{\ell}^{m}y^{\alpha}$ with $\ell=1,\dots,q-2$ are not even automatically orthogonal to each other since they carry the same eigenvalues of the Casimir and $J_3$. Thus, the orthonormalization of these states should be done by the usual Gram-Schmidt process on each pair of $y^{\alpha}\hat{Y}_{\ell}^{m},Y_{\ell}^{m}y^{\alpha}$ and then by the normalization on $y^{\alpha}$ as well as $y^{\alpha}\hat{Y}^{m}_{q-1}$.

\section{Solutions of a class of massive Chern-Simons equation of motion}
\label{AppendixCS}
We here consider a more generic EOM of a 3d massive Chern-Simons theory, which is useful for the diagonalization of several
of our blocks. Suppose we have the EOM
    \begin{equation}\label{massive CS EOM}
		\left(c\epsilon_{\mu\rho\nu}\partial^{\rho}-\frac{d}{z}\eta_{\mu\nu}  \right)A^{\nu}-i\left(\frac{1}{z^{1/2}}\partial_{\mu}+\frac{1}{2z^{3/2}}\delta^{z}_{\mu}\right)Y=0,
	\end{equation}
    where $\mu=0,1,z$.  Going to momentum space in the longitudinal directions $a=0,1$ by introducing momenta $k^0,k^1$
a possible solutions reads (with $ik=\sqrt{-k_ak^a}$)
	\begin{equation}
             \label{d2}
		\begin{aligned}
			&A^a=-\frac{1}{2k}\epsilon^{ab}k_{b}f K_{\frac{d}{c}}(ikz),\\
			&A^z=\frac{1}{4}f\left(K_{\frac{d}{c}-1}(ikz)-K_{\frac{d}{c}+1}(ikz)\right),\\
			&Y=\frac{c\sqrt{z}}{4i}f\left(K_{\frac{d}{c}-1}(ikz)+K_{\frac{d}{c}+1}(ikz)\right),
		\end{aligned}
	\end{equation}
    where $f\propto e^{-ik_ax^a}$ should be determined by  normalization conditions and $K_{\nu}(x)$ is the modified Bessel function. 
One can easily verify this solution by making us of 
    the recurrence relation of the modified Bessel function
	\begin{equation}
        \label{C3}
		x\partial_x K_{\nu}(x)=\pm\nu K_{\nu}(x)-xK_{\nu\pm1}(x),
	\end{equation}
	\begin{equation}
        \label{C4}
		x\partial_x K_{\nu\pm1}(x)=\mp(\nu\pm1)K_{\nu\pm1}(x)-xK_{\nu}(x).
	\end{equation}
 There exists (as usual) also a solution given in terms of modified Bessel functions of type $I_{\nu}(x)$ which we give below. For the purpose of 
  determining the spectrum we only need to worry about about one of these solutions as we can read off the mass parameter from the index of either
  of the Bessel functions. In order to construct the propagator we of course need both solutions, cf.\ eqn.~(\ref{eq:PropagatorBessel}).	

    In order to identify  the quantum fluctuations with a specific value of $\nu$ we organize $A^z, Y$ into linear combinations given in terms of a single Bessel function 
	\begin{equation}
			X^+ = f K_{\frac{d}{c}-1}(ikz),\quad
			X^- =-fK_{\frac{d}{c}+1}(ikz),
	\end{equation}
	where
	\begin{equation}
		X^{\pm}=2A^z\pm \frac{2i}{c\sqrt{z}}Y.
	\end{equation}
	 One has another set of independent solutions also expressed entirely in terms of $K_{\nu}$'s. 
       \begin{equation}
    \label{d7}
    \begin{aligned}
       A^{a}&=-\frac{1}{k}\eta^{ab}k_bfK_{\frac{d}{c}}(ikz),\\
        A^{z}&=\frac{1}{4}f(K_{\frac{d}{c}-1}(ikz)+K_{\frac{d}{c}+1}(ikz)),\\
        Y&=\frac{c \sqrt{z}}{4 i}f(K_{\frac{d}{c}-1}(ikz)-K_{\frac{d}{c}+1}(ikz)).
    \end{aligned}
    \end{equation}
We eliminate these by imposing the gauge fixing condition
    \begin{equation}
        \partial_a A^a=0. \label{gaugefixing}
    \end{equation}
    For completeness, let us also give the solution of eq.~(\ref{massive CS EOM}) in terms of $I_{\nu}(x)$ which differs slightly in the signs of some of the terms involved,
\begin{equation}
             \label{d2}
		\begin{aligned}
			&A^a=\frac{1}{2k}\epsilon^{ab}k_{b}g (I_{\frac{d}{c}}(ikz)),\\
			&A^z=\frac{1}{4}g\left(I_{\frac{d}{c}-1}(ikz)-I_{\frac{d}{c}+1}(ikz)\right),\\
			&Y=\frac{c\sqrt{z}}{4i}g\left(I_{\frac{d}{c}-1}(ikz)+I_{\frac{d}{c}+1}(ikz)\right), 
		\end{aligned}
	\end{equation}
 where $g\propto e^{-ik_ax^a}$ also should be determined by normalization conditions. Above equations can be verified by making use of the recurrence relations of the modified Bessel functions of type $I_{\nu}(x)$
  \begin{equation}
        \label{C5}
		x\partial_x I_{\nu}(x)=\pm\nu I_{\nu}(x)+xI_{\nu\pm1}(x),
	\end{equation}
	\begin{equation}
        \label{C6}
		x\partial_x I_{\nu\pm1}(x)=\mp(\nu\pm1)I_{\nu\pm1}(x)+xI_{\nu}(x).
	\end{equation}
Also in this case there exists a related solution which can be eliminated by the gauge choice in eq.~(\ref{gaugefixing}), namely
 \begin{equation}
    \label{d7}
    \begin{aligned}
       A^{a}&=\frac{1}{k}\eta^{ab}k_bgI_{\frac{d}{c}}(ikz),\\
        A^{z}&=\frac{1}{4}g(I_{\frac{d}{c}-1}(ikz)+I_{\frac{d}{c}+1}(ikz)),\\
        Y&=\frac{c \sqrt{z}}{4 i}g(I_{\frac{d}{c}-1}(ikz)-I_{\frac{d}{c}+1}(ikz)).
    \end{aligned}
    \end{equation}

\bibliographystyle{JHEP}




\end{document}